\documentclass[fleqn]{vch-book}
\usepackage{amsmath}
\usepackage{amssymb}
\usepackage{makeidx}
    \makeindex
\usepackage{times}
\usepackage{tabularx}

\usepackage{epsfig}

\newcommand{\be}{\begin{equation}}
\newcommand{\ee}{\end{equation}} 

\begin{document}


\begin{center}
{\huge \bf Graph Theory and the Evolution of Autocatalytic Networks \\}
\end{center}

\begin{center}
Sanjay Jain$^{a,b,c,d,*}$ and
Sandeep Krishna$^{b,*}$ \\
\end{center}

$^a${\it Department of Physics and Astrophysics, University of Delhi,
Delhi 110007, India} \\
$^b${\it Centre for Theoretical Studies,
Indian Institute of Science, Bangalore 560 012, India}\\
$^c${\it Santa Fe Institute, 1399 Hyde Park Road, Santa Fe, NM 87501, USA}\\
$^d${\it Jawaharlal Nehru Centre for Advanced Scientific Research,
Bangalore 560 064, India}\\
$^*$ Emails: jain@physics.du.ac.in, sandeep@physics.iisc.ernet.in\\         

{\bf Abstract:} 
We give a self-contained introduction to the 
theory of directed graphs, leading up to the relationship between
the Perron-Frobenius eigenvectors of a graph and its autocatalytic sets.
Then we discuss a particular dynamical system on a fixed but 
arbitrary graph, that describes the population dynamics
of species whose interactions are determined by the graph.
The attractors of this dynamical system 
are described as a function of graph topology.
Finally we consider a dynamical system in which the
graph of interactions of the species coevolves with the
populations of the species. We show that this system exhibits
complex dynamics including self-organization of the network
by autocatalytic sets, growth of 
complexity and structure, and collapse of the network followed by
recoveries. We argue that a graph
theoretic classification of perturbations of the network is helpful
in predicting the future impact of a perturbation over short and medium
time scales.\\

\section{Introduction}
Studies of networks are useful at several different levels (for recent reviews see
\cite{AB,DM,Strogatz,Watts}). At one level
one is interested in describing the structure of natural and man-made networks
such as food webs in ecosystems, biochemical and neural networks in organisms,
networks of social interaction among agents in societies, and technological
networks like the internet, etc. 
A useful representation of a network is a graph (and its generalizations)
where the components of the network (which could be species, neurons,
agents, etc.) are represented by nodes, and their mutual interactions by
the links of the graph. Graph theory provides important tools to
capture various aspects of the network structure.

At a second level one wants to know
how the network structure of the system influences what happens in the
system. E.g., the food-web structure of an ecosystem affects the dynamics
of populations of the species, the network of human contacts influences
the spread of a contagious disease, etc. At this level of discussion the
network is typically taken to be static on the time scales of interest; the
prime concern is the dynamics of other variables on a network
with some particular type of (fixed) structure. Here dynamical systems
theory is a major tool, and network variables (like the adjacency matrix elements
of the underlying graph) appear as fixed parameters in the dynamics of
other system variables like population, etc.

At a third level one is interested in how networks themselves change with time.
Biochemical, neural, ecological, social and technological networks are not static,
but are products of evolution.
Moreover this evolution is quite complex
in real systems. Networks sometimes self-organize and grow in size and
complexity, and sometimes disintegrate. Their evolution is usually intertwined
with other system variables, e.g., a food-web influences populations of
species, and if a species goes extinct, the food-web changes.
Understanding the processes and mechanisms involved in the evolution of
complex networks is a major intellectual challenge.

A problem that illustrates all these levels is the problem of the origin of
life on earth. The simplest living structure that we know ---- a bacterial cell ----
is a complex collection of several thousand types of molecules interacting with each other
in a complex network of chemical interactions. The network may be described by a graph
in which the nodes represent the molecular types or molecular
species, and links connecting nodes represent chemical interactions between the molecular
species. By participating in specific chemical reactions each molecular species or node
plays a rather definite functional role in the organization of the cell: it permits
or creates certain specific processes or spatial structures. Note that the complex
chemical network of a cell is needed to produce the processes and structures that exist in it,
and conversely, the same processes and structures are essential for maintaining the
network and allowing it to evolve. If
we assume that life originated on earth about 3.5 to 3.8 billion years ago as suggested
by the microfossil evidence, then
about 4 billion years back there was neither such a complex network of interactions
nor such processes and structures existing anywhere on the earth. 
One of the puzzles of the origin of life on earth is:
how did the network and the processes and spatial structures bootstrap themselves into existence
when none was present ---- how did a chemical `organization' emerge with individual molecular species
playing definite roles in it?

A second puzzle concerns the highly `structured' nature of the organization. The molecules
appearing in cells are very special (a small subset in a very large space of possible 
molecules) and so is the graph that describes their interactions (a special kind of graph in the
very large space of graphs). The probability of such structures arising by pure chance is astronomically
small. If we assume that it was not an unlikely chance event that created life, we are
led to the question: what then are the mechanisms that can create highly structured or 
`ordered' organizations? A similar question is relevant for economic and social networks.

In order to address such questions in a mathematical model,
one is naturally led to dynamical systems in which 
the graph describing the network is also a dynamical variable, whose dynamics is coupled to
that of other variables such as the population of the molecular species.
Here we present a model with such a structure, which has been inspired by the work in refs. 
\cite{Dyson,FKP,BFF,Kauffman2,BS,FB}. The analysis of such dynamical systems is facilitated by
the development of some new tools in graph theory. Another purpose of this article is to
discuss some of these new tools. Together, the model and these tools
address the above two questions about the origin
of life, and provide partial answers.
The model exhibits a mechanism by which a chemical organization can emerge where none existed
through the formation of small {\it autocatalytic sets} of molecular species.
In the model we also observe a {\it self-organizing process} which results in the growth
of the initial autocatalytic set into a complex and highly structured chemical
organization in a short time.

In addition, the model also captures, in an analytically tractable form, several phenomena that
one associates with the evolution of other biological and social systems. These
include emergence of cooperation and interdependence in the system; crashes and
recoveries of the system as a whole; `core-shifts';
appearance of `keystone species'; etc.
We also argue that the juxtaposition of graph theory and dynamical systems
provides the possibility of formulating more precisely notions that are
important and useful in everyday language but otherwise difficult to
pin down. In particular we attempt to formulate the notion of
`innovation' in this dynamical system, and classify innovations
into categories according to their graph theoretic structure. It turns out
that different categories of innovation have different short and longer term
impact on the dynamics of the system.

This article is organized roughly according to the three kinds of network
studies indicated above. In section 2 we discuss aspects of graph theory
in a self-contained manner, reviewing older results as well as recent work.
Among other things we describe a relation
between topological properties of a graph (namely its autocatalytic sets)
and its algebraic properties 
(the structure of the eigenvectors of its adjacency matrix).
In section 3 we discuss a simple dynamical system
describing molecular population dynamics on a fixed interaction graph. Here
we show how structure of the graph influences the dynamics of
the system; in particular relating the nature of its attractors to
graph topology. Section 4 describes a model of graph evolution,
motivated by the origin of life problem. In section 5 we show that
the dynamics of this model exhibits self-organization and growth of
cooperation and structure in the network, with analytical estimates of
the time scales involved. Section 6 discusses the phenomena of crashes
and recoveries exhibited by the model. In this section we also formulate
a definition of innovation that seems appropriate for this model, and
discuss a hierarchy of different categories of innovation and the roles they play in the
ups and downs of the system. Finally, section 7 contains a discussion
of some limitations of the model, speculations regarding the origin of life problem
and possible future directions.

\section{Graph theory and autocatalytic sets}

\subsection{Directed graphs and their adjacency matrices}
A {\it directed graph} $G = G(S,L)$, often referred to in the sequel as simply 
a {\it graph}, is defined by a set $S$
of `nodes' and a set $L$ of `links' (or `arcs'), where each
link is an ordered pair of nodes \cite{Harary,BG}. 
It is convenient to label the set of nodes by integers, $S=\{ 1,2,\ldots ,s\}$
for a graph of $s$ nodes. An example of a graph is
given in Figure \ref{digraphfig}a where each node is represented by a small labeled
circle, and a link $(j,i)$ is represented by an arrow pointing from node $j$ to node $i$.
A graph with $s$ nodes
is completely specified by an $s\times s$
matrix, $C=(c_{ij})$, called the {\it adjacency matrix} of the graph, and vice versa.
The matrix element in the $i^{\rm th}$ row and $j^{\rm th}$ column of $C$,
$c_{ij}$, equals unity if $L$ contains a directed link 
$(j,i)$ (arrow pointing from node $j$ to
node $i$), and zero otherwise. 
(This convention differs from the usual one where $c_{ij}=1$ if and only if there is a link
from node $i$ to node $j$; our adjacency matrix is the transpose of the usual one.
We have chosen this convention because it is more natural in the context of the
dynamical system to be discussed in subsequent sections.)
Figure \ref{digraphfig}b shows the adjacency
matrix corresponding to the graph in Figure \ref{digraphfig}a.
We will use the terms `graph' and `adjacency matrix' interchangeably: the phrase
`a graph with adjacency matrix $C$' will often be abbreviated to `a graph $C$'.
Undirected graphs are special cases of directed graphs whose adjacency matrices
are symmetric. A single (undirected) link of an undirected graph between, say, nodes $j$ 
and $i$,
can be viewed as two directed links of a directed graph, one from $j$ to $i$ and the other
from $i$ to $j$.

A graph $G'=G'(S',L')$ is called a {\it subgraph} of $G(S,L)$ if $S' \subset S$ and $L' \subset L$.
We will use the term `subgraph' if $G'$ satisfies a stronger property: every link in
$L$ with both endpoints in $S'$ also belongs to $L'$. That is, for us, a subgraph will
be a subset of nodes together with {\it all} their mutual links. (This is often called
an `induced subgraph' in the literature \cite{BG}.)
The graph in Figure \ref{digraphfig}c (comprising nodes 14, 15, 16, 17, 18 and 19 and all their
mutual links) is thus a subgraph of the graph in
Figure \ref{digraphfig}a. For a subgraph we will often find it more convenient to label the nodes
not by integers starting from 1, but by the same labels the corresponding nodes had in the parent graph. 
The adjacency matrix of a subgraph can be obtained by deleting all the
rows and columns from the full adjacency matrix that correspond to the nodes
outside the subgraph. The highlighted portion of the matrix in Figure \ref{digraphfig}b is the
adjacency matrix of the subgraph in Figure \ref{digraphfig}c. 

A {\it walk} of length $n$ (from node $i_1$ to node $i_{n+1}$)
is an alternating sequence of nodes and links $i_1l_1i_2l_2\ldots 
i_nl_ni_{n+1}$ such that
link $l_1$ points from node $i_1$ to node $i_2$
(or $l_1 = (i_1, i_2)$), $l_2$ points from $i_2$ to $i_3$ and so on. 
 A walk with all nodes distinct 
(except possibly the first and last nodes) will be called
a {\it path}. If the first and last nodes $i_1$ and $i_{n+1}$
of a walk or path are the same, it will be referred to as a {\it closed} walk or path. 
The existence of even one closed walk
in the graph implies the existence of an infinite number of distinct walks in the graph.
In the graph of Figure \ref{digraphfig}a, there is an infinite number of walks from node 11 to node 17 
(e.g., $11\rightarrow 12\rightarrow 14\rightarrow 17$,
$11\rightarrow 12\rightarrow 11\rightarrow 12\rightarrow 14\rightarrow 17$, $\ldots$) 
but no walks from node 11 to node 10. 
An undirected graph trivially has closed walks if it has any undirected links at all.

In the graph theory literature, what we have defined above to be a `closed path' is usually
referred to as a `cycle'. However, for later convenience, we define a cycle somewhat 
differently. We define an {\it $n$-cycle} to be a
subgraph with $n \geq 1$ nodes which contains exactly $n$ links and also contains
a closed path that covers all
$n$ nodes. E.g., the subgraph formed by node 20 and its self link
is a 1-cycle, that formed by nodes 1 and 2 is a 2-cycle and by nodes 3,4 and 5 a 3-cycle. The
subgraph formed by nodes 1,2,3,4 and 5 is not a 5-cycle because it does not have a closed 
path covering all the five nodes. The word `cycle' will be used generically for an $n$-cycle of
unspecified length.

Given a directed graph $C$, its {\it associated undirected graph} (or `symmetrized version')
$C^{(s)}$ can be obtained by adding additional links as follows: for every link $(j,i)$ 
in $L$, add another link $(i,j)$ if the latter is not already in $L$. 
Two nodes of a directed graph $C$ will be said to be {\it connected} if there exists a path between
them in the associated undirected graph $C^{(s)}$, and disconnected otherwise.
Thus any directed graph can be decomposed into `connected components' which are maximal
sets of connected nodes
(e.g., the graph of Figure \ref{digraphfig}a has five connected components that are disconnected from each
other). In a directed graph $C$, 
we refer to a node $i$ as being `downstream' from a
node $j$ if there is a path in $C$ leading from $j$ to $i$, and no path from $i$ to $j$.
Similarly $i$ is `upstream'
from $j$ if there is a path in $C$ leading from $i$ to $j$,
and no path from $j$ to $i$. Thus in Figure \ref{digraphfig}a,
node 17 is downstream from node 11, or equivalently node 11 is upstream from
node 17. Node 10 is neither upstream nor downstream from node 11
since they are not connected, and node 12 is neither upstream nor downstream
from 11 because each can be reached from the other along some directed path.

If $C$ is the adjacency matrix of a graph then it is easy to see that
$(C^n)_{ij}$ equals the number of distinct walks of length $n$ from node
$j$ to node $i$. E.g., $C^2_{ij} = \sum_{k=1}^s C_{ik}C_{kj}$; each term in the
sum is unity if and only if there exists a link from $j$ to $k$ and from $k$ to $i$;
hence the sum counts the number of walks from $j$ to $i$ of length 2.
\\

\noindent
{\bf Perron-Frobenius eigenvalues and eigenvectors (PFEs)}\\
A vector ${\bf x}=(x_1,x_2,\ldots,x_s)$ is said to be an eigenvector of
an $s\times s$ matrix $C$ with an eigenvalue $\lambda$ if 
for each $i$, 
$\sum_{j=1}^{s} c_{ij}x_j=\lambda x_i$.
The eigenvalues of a matrix $C$ are roots of the {\it characteristic
equation} of the matrix: $|C-\lambda I|=0$
where $I$ is the identity matrix of the same dimensionality as $C$ and $|A|$ is the determinant of the
matrix $A$. In general a matrix will have complex eigenvalues and eigenvectors,
but an adjacency matrix of a graph has special properties, because it is a `non-negative'
matrix, i.e., it has no negative entries.

For any non-negative matrix, the
Perron-Frobenius theorem \cite{Seneta,BP} guarantees that there exists an eigenvalue which is
real and larger than or equal to all other eigenvalues in magnitude. This
largest eigenvalue is often called the Perron-Frobenius eigenvalue of the matrix, which
we will denote by $\lambda_1(C)$ for a graph $C$.
Further the theorem also states that there exists an eigenvector of $C$
corresponding to $\lambda_1(C)$ (which we will refer to as a Perron-Frobenius Eigenvector, PFE)
all of whose components are
real and non-negative. 
The Perron-Frobenius eigenvalue of the graph in Figure \ref{digraphfig}a is 1. Four 
PFEs of the graph in Figure \ref{digraphfig}a are displayed in Figure \ref{digraphfig}d.

The presence or absence of closed paths in a graph can be determined from the
Perron-Frobenius eigenvalue of its adjacency matrix (see ref. \cite{JK2}
for a simple proof):\\

\noindent \underline{Proposition 1.}  
{\it If a graph, $C$,} \\
(i)  {\it has no closed walk then $\lambda_1(C)=0$},\\
(ii) {\it has a closed walk then $\lambda_1(C) \geq 1$},\\
(iii) {\it has a closed walk and all closed walks only occur in subgraphs that are
cycles then $\lambda_1(C) = 1$}.\\
Note that $\lambda_1$ cannot take values between zero and one because of the discreteness
of the entries of $C$ which are either zero or one.
(Thus, for an undirected graph, if it has even one undirected link, $\lambda_1(C) \geq 1$.)
Several results pertaining to the relationship of the graph structure to the
structure of its PFEs can be found in ref. \cite{Rothblum}.\\

\noindent
{\bf Irreducible graphs and matrices}\\
A subgraph of a directed graph is termed {\it irreducible}
if there is a path within the subgraph from each node in the subgraph to every other node in 
the subgraph. The simplest irreducible subgraph is a 1-cycle.
In Figure \ref{digraphfig}a
the subgraph comprising nodes 3,4 and 5 is irreducible, as is the subgraph of nodes 6 and 7,
but the subgraph of nodes 3,4,5,6 and 7 is not irreducible since there
is, for example, no path from node 6 to node 5.

If a graph or subgraph is irreducible then the corresponding adjacency matrix is also
termed {\it irreducible}. Thus a matrix $C$ is {\it irreducible} if for every
ordered pair of nodes $i$ and $j$ there exists a positive integer $k$ such that
$(C^k)_{ij}>0$. Refs. \cite{Seneta,BP} describes further properties of irreducible matrices.\\

The nodes of any graph can be grouped into a unique set of irreducible subgraphs
as follows:\\
(1) Pick any node, say $i$. Find all the nodes which have paths leading to
them starting at $i$. Denote this set by $S_1$; it may include $i$ itself. Similarly find all the nodes
which have paths leading to $i$. Denote this set by $S_2$. Denote the subgraph
formed by the set of nodes $\{i\}\cup (S_1\cap S_2)$ and all their mutual links as
$C_1$. If $S_1\cap S_2 \neq \Phi$ ($\Phi$ denotes the empty set), then $C_1$ is an irreducible graph
because every node of $C_1$ has a path within $C_1$ to every other node in it.
If $S_1\cap S_2 = \Phi$,
then $i$ does not belong to any irreducible subgraph and $C_1$ consists of just the node $i$ and no 
links. \\
(2) Pick another node which is not in $C_1$ and
repeat the procedure with that node to get another subgraph, $C_2$. The
sets of nodes comprising the two subgraphs will be disjoint.\\
(3) Repeat this process until all nodes have been placed in some $C_{\alpha}$, 
$\alpha = 1,2, \ldots, M$. Each $C_{\alpha}$ is either an irreducible subgraph or
consists of a single node with no links.

Irrespective of which nodes are picked and in which order,  
this procedure will produce for any graph
a unique set of disjoint subgraphs (upto labelling of
the $C_{\alpha}$) encompassing all the nodes of the graph.
The graph in Figure \ref{digraphfig}a will decompose into 14 such subgraphs (see Figure \ref{digraphfig}e).

We say there is  a path from an irreducible subgraph $C_1$ to another irreducible
subgraph $C_2$ if there is a path in $C$ from any node of $C_1$ to any node of $C_2$.
The terms `downstream' and `upstream' can thus be used unambiguously for
the $C_{\alpha}$.\\

\noindent
{\bf Decomposition of a general graph}\\
A general adjacency matrix can be rewritten in a useful form by renumbering
the nodes by the following procedure \cite{Seneta,BP}:

\noindent
Determine all the subgraphs $C_1, C_2, \ldots, C_M$ of the graph as described above.
Construct a new graph of $M$ nodes, one node for each $C_{\alpha}$, $\alpha = 1,\ldots,M$.
The new graph has a directed link from $C_{\beta}$ to $C_{\alpha}$ if, in the original graph, any node of
$C_{\beta}$ has a link to any node of $C_{\alpha}$. Figure \ref{digraphfig}e illustrates
what this new graph looks like for the graph of Figure \ref{digraphfig}a.

Clearly the resulting graph cannot have any closed paths. For if it were to have a closed path then
the $C_{\alpha}$ subgraphs comprising the closed path would together have formed a larger
irreducible subgraph in the first place. Therefore we can renumber the $C_{\alpha}$
such that if $\alpha > \beta$, $C_{\beta}$ is never downstream from $C_{\alpha}$.
Now we can renumber the nodes of the original graph such that nodes belonging to
a given $C_{\alpha}$ occupy contiguous node numbers, and whenever a pair of nodes
$i$ and $j$ belong to different subgraphs $C_{\alpha}$ and $C_{\beta}$ respectively,
then $\alpha > \beta$ implies $i > j$. Such a renumbering is in general not unique,
but with any such renumbering the 
adjacency matrix takes the following canonical form:

$$C=\left(\begin{array}{cccccc}
C_1 &&&&& 0 \\
& C_2 &&&&\\
&& . &&&\\
&&& . &&\\
&&&&. &\\
R &&&&& C_M\\
\end{array}\right)$$
where $0$ indicates that the upper block triangular part of the matrix contains
only zeroes while the lower block triangular part, $R$, is not equal to zero in general.
It can be seen that the graph in Figure \ref{digraphfig}a is already in this canonical form.
In Figure \ref{digraphfig}b, the dotted lines demarcate the block diagonal portions which
correspond to the $C_{\alpha}$.

From the above form of $C$ it follows that
$$ |C-\lambda I|=|C_1-\lambda I|\times |C_2-\lambda I|\times\ldots\times|C_M-\lambda I|$$
Therefore the set of eigenvalues of $C$ is the union of the sets of eigenvalues of
$C_1,\ldots,C_M$. $\lambda_1(C)={\rm max}_{\alpha}\{\lambda_1(C_{\alpha})\}$.

Therefore if a given graph $C$ has a Perron-Frobenius eigenvalue $\lambda_1 > 0$ then
it contains at least one irreducible subgraph with Perron-Frobenius eigenvalue
$\lambda_1$. When $\lambda_1 > 0$, 
all irreducible subgraphs of $C$ with Perron-Frobenius eigenvalue
equal to $\lambda_1$ are referred to as {\it basic} subgraphs. The yellow nodes in
Figure \ref{digraphfig}e correspond to the basic subgraphs of Figure \ref{digraphfig}a.\\

\noindent
\subsection{Autocatalytic sets}
The concept of an autocatalytic set (ACS) was first introduced in the context of a set of
catalytically interacting molecules. There it was defined to be a set of
molecular species which contains a catalyst for each of its member species
\cite{Eigen,Kauffman1,Rossler}. Such a set of molecular species
can collectively self-replicate under certain circumstances even if none of its component molecular
species can individually self-replicate. This property is considered important in understanding
the origin of life. If we imagine a node in a directed graph to represent a molecular
species and a link from $j$ to $i$ as signifying that $j$ is a catalyst for $i$, this motivates
the following graph-theoretic definition of an ACS in any directed graph:
An {\it autocatalytic set} (ACS) is a
subgraph, each of whose nodes
has at least one incoming link from a node belonging to the
same subgraph.

Figure \ref{acsfig} shows various ACSs. The simplest ACS is a 1-cycle; Figure \ref{acsfig}a. 
There is the following hierarchical relationship between cycles, irreducible subgraphs and ACSs:
all cycles are irreducible subgraphs and all irreducible subgraphs are ACSs, but not all ACSs are
irreducible subgraphs and not all irreducible subgraphs are cycles.
Figures \ref{acsfig}a and \ref{acsfig}b are graphs that are irreducible as well as cycles, 
\ref{acsfig}c is an ACS that is not an irreducible subgraph and hence not a cycle,
while \ref{acsfig}d and \ref{acsfig}e are examples of irreducible graphs that are not cycles.
It is not difficult to see the following \cite{JK2}:\\

\noindent \underline{Proposition 2.}\\
(i) {\it An ACS must contain a closed path}. Consequently,\\
(ii) {\it If a graph $C$ has no ACS then $\lambda_1(C)=0$}.\\
(iii) {\it If a graph $C$ has an ACS then $\lambda_1(C) \geq 1$}.\\

\noindent
{\bf Relationship between autocatalytic sets and Perron-Frobenius eigenvectors}\\
The ACS is a useful graph-theoretic construct in part because of its connection with the PFE.
Let ${\bf x}$ be a PFE of a graph. Consider the set of all 
nodes $i$ for which $x_i$ is
non-zero. We will call the subgraph of all these nodes and their mutual links 
the `subgraph of the PFE {\bf x}'. If all the components
of the PFE are non-zero then the subgraph of the PFE is the entire graph.
For example the subgraph of the PFE ${\bf e}_3$ mentioned in Figure \ref{digraphfig}d is 
is the graph shown in Figure \ref{digraphfig}c. One can show that \cite{JK2} \\

\noindent \underline{Proposition 3}\\
{\it If $\lambda_1(C) > 0$, then the subgraph of any PFE of $C$ is an ACS}.\\

For the PFEs of Figure \ref{digraphfig}d this is immediately verified by inspection.
Note that this result relates an algebraic property of a graph, its PFE, to
a topological structure, an ACS. Further, this result is not true if we considered
irreducible graphs intead of ACSs. E.g., the subgraph of ${\bf e}_3$, shown in 
Figure \ref{digraphfig}c, is not an irreducible graph.

Note also that the converse of the above statement is not true, 
i.e., there need not exist a PFE for every ACS
in a given graph. Thus in Figure \ref{digraphfig}a, nodes 3,4,5,6 and 7 form an ACS but there is no
eigenvector with eigenvalue $\lambda_1$ for which all these and only these components 
are non-zero.

Let ${\bf x}$ be a PFE of a graph $C$, and let 
$C'$ denote the adjacency matrix of the subgraph of ${\bf x}$. Let
$\lambda_1(C')$ denote the
Perron-Frobenius eigenvalue of $C'$. It is not difficult to see that
$\lambda_1(C') = \lambda_1(C)$.
Figure \ref{PFEfig} illustrates this point. For the graph in Figure \ref{PFEfig}a
$\lambda_1=1$. Figure \ref{PFEfig}b shows a PFE of the graph and how it satisfies the 
eigenvalue equations. For this PFE, nodes 1, 5 and 6 have $x_i=0$. Removing
these nodes produces the PFE subgraph shown in Figure \ref{PFEfig}c. Its adjacency matrix, $C'$,
is obtained by removing rows 1, 5, 6 and columns 1, 5, 6 from the original
matrix. Figure \ref{PFEfig}d illustrates that the vector constructed by removing
the zero components of the PFE is an eigenvector of $C'$ with eigenvalue 1. 
The logic of this example is easily extended to a general proof that 
$\lambda_1(C') = \lambda_1(C)$. 

We can now perform a graph decomposition of $C'$ into irreducible subgraphs as before; 
since $\lambda_1(C') = \lambda_1(C)$, it follows that 
$C'$ must contain at least one of the basic subgraphs of $C$.
If $C'$ contains only one of the basic subgraphs of $C$ we will refer to ${\bf x}$ as a
{\it simple} PFE, and to $C'$ as a {\it simple ACS}. The graph in Figure \ref{digraphfig}a has only four
simple PFEs which are displayed in Figure \ref{digraphfig}d. All PFEs of $C$ are linear combinations of
its simple PFEs.\\

\noindent
{\bf Core and periphery of a simple PFE}\\
If $C'$ is the subgraph of a simple PFE, 
the basic subgraph of $C$ contained in $C'$ will be called the {\it core} of $C'$
(or equivalently, the `core of the simple PFE'), and denoted $Q'$.
The set of the remaining nodes and links of $C'$ that are not in its core 
will together be said to constitute the {\it periphery} of $C'$.
For example, for the PFE in Figure \ref{digraphfig}c the core is the 2-cycle comprising nodes 14 and 15.
Note that the periphery is not a subgraph in the sense we are using the word `subgraph', 
since it contains links not just between periphery nodes but also from nodes outside the 
periphery (like the link from node 15
to 16 in Figure \ref{digraphfig}c).

The core and periphery can be shown to have the following topological property
(which justifies the nomenclature):\\

\noindent \underline{Proposition 4.} {\it 
From every node in the core of (the subgraph of) a simple PFE there exists
a path leading to every other node of the PFE subgraph. From no periphery node
is there any path leading to any core node.}\\ 
Thus all periphery nodes are
downstream from all core nodes.
Starting from the core one can reach the periphery but not vice versa.\\

It follows from the Perron-Frobenius theorem for irreducible graphs \cite{Seneta}
that $\lambda_1(Q')$ will necessarily increase if any link is added to the core.
Similarly removing any link will decrease $\lambda_1(Q')$.
Thus $\lambda_1$ measures the
multiplicity
of internal pathways in the core. Figure \ref{intpathfig} illustrates this point.
\\

\noindent
{\bf Core and periphery of a non-simple PFE}\\
Since any PFE of a graph can be written as a linear combination of
a set of simple PFEs (this set is unique for any graph),
the definitions of core and periphery can be readily extended to any PFE
as follows:\\
The {\it core of a PFE}, denoted $Q$, is the union of the cores of those simple 
PFEs whose linear combination forms the given PFE.
The rest of the nodes and links of the PFE subgraph constitute its periphery.
It follows from the above discussion
that $\lambda_1(Q)=\lambda_1(C)$.
When the core is a union of disjoint cycles 
then $\lambda_1(Q) = 1$, and
vice versa.\\

\noindent
{\bf The structure of PFEs when there is no ACS}\\
The above discussion about the structure of PFEs 
was for graphs $C$ with $\lambda_1(C) > 0$. If $\lambda_1(C) = 0$, the
graph has no ACS. Then the structure of PFEs is as follows: there exists a PFE for
every connected component of the graph. Since there
are no closed walks in the graph, all walks have
finite lengths. Consider the longest paths in a given connected component. Identify
the nodes that are the endpoints of these longest paths. The PFE corresponding to the given
connected
component will have $x_i > 0$ for each of the latter nodes and $x_i = 0$ for all
other nodes in the graph. Again a general PFE is a linear combination of all such
PFEs, one for each connected component of the graph. In this case since there 
is no closed path there is no core (or periphery) for any PFE 
of the graph. The core of all PFEs of such
a graph may be defined to be the null set, $Q = \Phi$.\\

\section{A dynamical system on a fixed graph}

In the previous section we have discussed the properties of graphs and their
associated adjacency matrices, eigenvalues and eigenvectors. In this section
we discuss the dynamical significance of the same constructs. In particular,
we present an example of a dynamical system on a fixed graph described by
a set of coupled ordinary differential equations, whose attractors are
precisely the PFEs discussed above. This dynamical system arises as an
idealization of population dynamics of a set of chemicals.

Consider the simplex of normalized non-negative vectors in $s$ dimensions:
$J = \{ {\bf x} \equiv (x_1,x_2,\ldots,x_s) \in {\bf R}^s |
0 \leq x_i \leq 1, \sum_{i=1}^s x_i = 1\} $.
For a fixed graph $C = (c_{ij})$ with $s$ nodes, consider the set of
coupled ordinary differential equations \cite{JK1}
\begin{equation}
\dot{x}_i=\sum_{i=1}^s c_{ij}x_j-x_i\sum_{j,k=1}^{s} c_{kj}x_j.
\label{xdot}
\end{equation}
This will be the dynamical system of interest to us in this section.

Note that the dynamics preserves the normalization of ${\bf x}$,
$\sum_{i=1}^s \dot{x}_i = 0$. For non-negative $C$ it leaves the simplex
$J$ invariant. (For negative $c_{ij}$, additional conditions have
to be added (see \cite{JK3}) but we do not discuss that case here.)

The links of the graph
represent the interactions between the variables $x_i$ that
live on the nodes. $x_i$ could represent, for example, 
the relative population of the $i^{\rm th}$ species in a population
of $s$ species, or the probability of the $i^{\rm th}$ strategy 
among a group of $s$ strategies in an evolutionary game, or the
market share of the $i^{\rm th}$ company among a set of competing
companies, etc. It is useful to see how equation (\ref{xdot}) arises in a 
population dynamic context.

Let $i\in \{1,\ldots,s\}$ denote a chemical (or molecular) species in a
chemical reactor. Molecules can react with each other in various ways; we focus
on only one aspect of their interactions: catalysis.
The catalytic interactions can
be described by a directed graph with $s$ nodes. The nodes represent the $s$
species and the existence of a link from node $j$ to node $i$ means that
species $j$ is a catalyst
for the production of species $i$.
In terms of the adjacency matrix, $C=\{c_{ij}\}$
of this graph, $c_{ij}$ is set to unity if $j$ is a catalyst of $i$ and
is set to zero otherwise. The operational meaning of catalysis is as follows:

Each species $i$ will have an associated non-negative population $y_i$ in the pond
which changes with time. In a certain approximation (discussed below) the
population dynamics for a fixed set of chemical species whose interactions
are given by $C$, will be given by
\begin{equation}
\dot{y}_i=\sum_{j=1}^s c_{ij}y_j -\phi y_i,
\label{ydot}
\end{equation}
where $\phi(t)$ is some function of time. To see how such an equation might
arise,
assume that species $j$ catalyses the ligation of reactants $A$ and $B$ to
form the species $i$, $A+B\stackrel{j}{\rightarrow} i$. Then the rate of growth
of the population $y_i$ of species $i$ in a well stirred reactor will be given
by $\dot{y_i}=k(1+\nu y_j)n_An_B-\phi y_i$, where $n_A,n_B$ are reactant
concentrations, $k$ is the rate constant for the spontaneous reaction, $\nu$ is
the catalytic efficiency, and $\phi$ represents a common death rate or dilution
flux in the reactor. Assuming the catalysed reaction is much faster than the
spontaneous reaction, and that the concentrations of the reactants are large
and fixed, the rate equation becomes $\dot{y_i}=Ky_j-\phi y_i$, where $K$ is a
constant.
In general since species $i$ can have multiple catalysts, we get
$\dot{y}_i=\sum_{j=1}^s K_{ij}y_j -\phi y_i$, with $K_{ij}\sim c_{ij}$.
We make the further idealization $K_{ij}=c_{ij}$ giving equation (\ref{ydot}).

The relative population of species $i$ is by definition
$x_i \equiv y_i/\sum_{j=1}^s y_j$. Therefore 
${\bf x} \equiv (x_1,\ldots,x_s) \in J$, since
$0\le x_i\le 1, \sum_{i=1}^sx_i=1$. Taking the time derivative of
$x_i$ and using (\ref{ydot}) 
it is easy to see that $\dot{x}_i$ is given by (\ref{xdot}).
Note that the $\phi$ term, present in (\ref{ydot}), cancels out and is
absent in (\ref{xdot}).

We remark that the quasispecies equation \cite{Eigen} has the same form as equation (\ref{ydot}),
albeit with a different interpretation and a special structure of the $C$ matrix
that arises from that interpretation.\\

\subsection{Attractors of equation (\ref{xdot})}
The rest of this section consists of examples and arguments to justify the

\noindent
\underline{Proposition 5.} {\it For any graph $C$, }\\
(i) {\it Every eigenvector of $C$ that belongs to $J$ is a 
fixed point of (\ref{xdot}), and vice versa.} \\
(ii) {\it Starting from any initial condition in the simplex $J$, the
trajectory converges to some fixed point (generically denoted 
${\bf X}$) in $J$.} \\ 
(iii) {\it For generic initial conditions in $J$, ${\bf X}$ is a
Perron-Frobenius eigenvector (PFE) of $C$.} (For special initial
conditions, forming a space of measure zero in $J$, ${\bf X}$ could
be some other eigenvector of $C$. Henceforth we ignore such special
initial conditions.) \\
(iv) {\it If $C$ has a unique PFE, ${\bf X}$ is the unique stable attractor
of (\ref{xdot}). } \\
(v) {\it If $C$ has more than one linearly independent PFE, then ${\bf X}$
can depend upon the initial conditions. The set of allowed ${\bf X}$ is
a convex linear combination of a subset of the PFEs.} 
The interior of this convex set in
$J$ may then be said to be the `attractor' of (\ref{xdot}), in the sense
that for generic initial conditions all trajectories
converge to a point in this set. \\
(vi) {\it For every ${\bf X}$ belonging to the attractor set, the set
of nodes $i$ for which $X_i > 0$ is the same and is uniquely determined
by $C$. The subgraph formed by this set of nodes will be called the 
`subgraph of the attractor' of (\ref{xdot}) for the graph $C$.}
Physically, this set consists of nodes that always end up with a nonzero
relative population when the dynamics (\ref{xdot}) is allowed to
run its course, starting from generic initial conditions. \\
(vii) {\it If $\lambda_1(C) > 0$, the subgraph of the attractor of (\ref{xdot})
is an ACS.} This ACS will be called 
the {\it dominant ACS} of the graph. The dominant ACS is independent of
(generic) initial conditions and depends only on $C$.

For example for the graph of Figure \ref{digraphfig}a, ${\bf X}$ is a convex linear
combination of ${\bf e}_2$ and ${\bf e}_3$, ${\bf X} = a{\bf e}_2 
+ (1-a){\bf e}_3$, with $0 \leq a \leq 1$. $a$ depends upon initial
conditions; generically $0 < a < 1$. The subgraph of the attractor 
contains eight nodes, 6,7,14-19. Starting with generic initial conditions
where all the $x_i$ are nonzero, the trajectory will converge to a point
${\bf X}$ where these eight nodes have nonzero $X_i$ and each of the
other twelve nodes have $X_i =0$. The eight populated nodes
form an ACS, the dominant ACS of the graph.

To see (i), let ${\bf x}^{\lambda} \in J$ be an eigenvector of $C$, 
$\sum_j c_{ij} x_j = \lambda x_i$. Substituting this on the r.h.s. of 
(\ref{xdot}), one gets zero. Conversely, if the r.h.s. of (\ref{xdot})
is zero, one finds ${\bf x} = {\bf x}^{\lambda}$, with $\lambda = 
\sum_{k,j} c_{kj} x_j$. 

To motivate (ii) and (iii) it is most convenient
to consider the underlying dynamics (\ref{ydot}) from which
(\ref{xdot}) is derived:
Since (\ref{xdot}) is independent of $\phi$, we can set $\phi=0$ in
(\ref{ydot}) without any loss of generality. With $\phi=0$ the general
solution of (\ref{ydot}), which is a linear system,
can be schematically written as:
$${\bf y}(t)=e^{Ct}{\bf y}(0),$$
where ${\bf y}(0)$ and ${\bf y}(t)$ are viewed as column vectors.
Suppose ${\bf y}(0)$ is a right eigenvector of $C$ with eigenvalue $\lambda$,
denoted ${\bf y}^\lambda$. Then
$${\bf y}(t)=e^{\lambda t}{\bf y}^{\lambda}.$$
Since this time dependence is merely a rescaling of the eigenvector,
this is an alternative way of seeing that 
${\bf x}^{\lambda}={\bf y}^{\lambda}/\sum_{j=1}^s y_j^{\lambda}$
is a fixed point of (\ref{xdot}).
If the eigenvectors of $C$ form a basis in $R^s$, 
${\bf y}(0)$ is a 
linear combination: ${\bf y}(0) = \sum_{\lambda} a_{\lambda}{\bf y}^{\lambda}$.
In that case, for
large $t$ it is clear that the term with the largest value of $\lambda$ will
win out, hence
$${\bf y}(t)\stackrel{t\rightarrow\infty}{\sim} e^{\lambda_1 t}{\bf y}^{\lambda_1}$$
where $\lambda_1$ is the eigenvalue of $C$ with the largest real part (which
we know is the same as its Perron-Frobenius eigenvalue) and ${\bf y}^{\lambda_1}$ an 
associated eigenvector. Therefore, for generic initial conditions
the trajectory of (\ref{xdot}) will converge to 
${\bf X} = {\bf x}^{\lambda_1}$, a PFE of $C$. If the eigenvectors of $C$ do not
form a basis in $R^s$, the above result is still true (as we will see in examples). 

Note that
$\lambda_1$ can be interpreted as the `population growth rate' at large $t$,
since
$\dot{\bf y}(t)\stackrel{t\rightarrow\infty}{\sim} \lambda_1 {\bf y}$.
In the previous section we had mentioned that $\lambda_1$ measures a
topological property of the graph, namely, the multiplicity of internal
pathways in the core of the graph. Thus in the present model, 
$\lambda_1$ has both a topological and dynamical significance, which
relates two distinct properties of the system, 
one structural (multiplicity of pathways in the core of the graph), and the other dynamical
(population growth rate). The higher the multiplicity of pathways in the core, the
greater is the population growth rate of the dominant ACS.

Part (iv) follows from the above. We will give examples as illustrations of (v) and (vi).
Further, from Proposition 3, previous section, we 
know that the subgraph of a PFE has to be an ACS, whenever $\lambda_1 > 0$.
That explains (vii).
It is instructive to consider examples of graphs and see how the 
trajectory converges to a PFE.\\

\noindent
{\bf Example 1}. A simple chain, Figure \ref{uniquePFEfig}a:\\
The adjacency matrix of this graph has all eigenvalues (including $\lambda_1$)
zero. There is only one (normalized) eigenvector corresponding to this eigenvalue,
namely ${\bf e} = (0,0,1)$ and this is the unique PFE of the graph. (This
is an example where the eigenvectors of $C$ do not form a basis in $R^s$.)
Since node 1 has no catalyst, its rate equation is (henceforth taking $\phi =0$)
$\dot{y_1}=0$. Therefore $y_1(t)=y_1(0)$, a constant. The
rate equation for node 2 is $\dot{y_2}=y_1=y_1(0)$. Thus $y_2(t)=y_2(0)+y_1(0) t$.
Similarly $\dot{y_3} = y_2$ implies that $y_3(t) = (1/2) y_1(0) t^2 + y_2(0) t + y_3(0)$.
At large $t$, $y_1 = {\rm constant}$, $y_2 \sim t$, $y_3 \sim t^2$; hence $y_3$
dominates. Therefore, $X_i = \lim_{t \rightarrow \infty} x_i(t)$ is given by
$X_1=0,X_2=0,X_3=1$. Thus we find that ${\bf X}$ equals the unique PFE ${\bf e}$,
independent of initial conditions.  \\

\noindent
{\bf Example 2}. A 1-cycle, Figure \ref{uniquePFEfig}b:\\
This graph has two eigenvalues, $\lambda_1=1$,
$\lambda_2=0$. The unique PFE is ${\bf e}=(1,0)$. The rate equations are 
$\dot{y_1}=y_1, \dot{y_2}=0$, with the solutions $y_1(t) = y_1(0)e^t, y_2(t)=y_2(0)$.
At large $t$ node 1 dominates, hence ${\bf X}=(1,0)={\bf e}$. The exponentially 
growing population of 1 is a consequence of the fact that 1 is a self-replicator,
as embodied in the equation $\dot{y_1}=y_1$. \\

\noindent
{\bf Example 3}. A 2-cycle, Figure \ref{uniquePFEfig}c:\\
The corresponding adjacency matrix has
eigenvalues $\lambda_1=1, \lambda_2 = -1$. The unique normalized PFE is 
${\bf e}=(1/2, 1/2)$. 
The population dynamics equations are $\dot{y_1}=y_2, \dot{y_2}=y_1.$
The general solution to these is (note $\ddot{y_1}=y_1$)
$$y_1(t) = Ae^t + Be^{-t}, \quad \quad y_2(t) = Ae^t - Be^{-t}.$$
Therefore at large $t$, $y_1 \rightarrow Ae^t, y_2 \rightarrow Ae^t$, hence
${\bf X}=(1,1)/2 = {\bf e}$. Neither 1 nor 2 is individually a self-replicating
species, but collectively they function as a self-replicating entity. This
is true of all ACSs. \\

\noindent
{\bf Example 4}. A 2-cycle with a periphery, Figure \ref{uniquePFEfig}d:\\
This graph has $\lambda_1=1$ and a unique normalized PFE ${\bf e}=(1,1,1)/3$.
The population equations for $y_1$ and $y_2$ and consequently their general solutions
are the same as Example 3, but now in addition $\dot{y_3}=y_2$, yielding
$y_3(t)=Ae^t + Be^{-t} + {\rm constant}$. Again for large $t$, $y_1,y_2,y_3$ 
grow as $\sim Ae^t$, hence ${\bf X} = (1,1,1)/3 = {\bf e}$. The dominant
ACS includes all the three nodes.

This example shows how a parasitic periphery (which does not feed back into
the core) is supported by an autocatalytic core. This is also an example of
the following general result: when a subgraph $C'$, with largest eigenvalue
$\lambda_1'$, is {\it downstream} from another subgraph $C''$ with largest
eigenvalue $\lambda_1'' > \lambda_1'$, then the population of the former
also increases at the rate $\lambda_1''$. Therefore if $C''$ is populated in
the attractor, so is $C'$. In this example $C'$ is the single node 3 with
$\lambda_1'=0$ and $C''$ is the 2-cycle of nodes 1 and 2 with $\lambda_1''=1$. \\

\noindent
{\bf Example 5}. A 2-cycle and a chain, Figure \ref{uniquePFEfig}e: \\
The graph in Figure \ref{uniquePFEfig}e combines the graphs of Figures \ref{uniquePFEfig}a and c.
Following the analysis of those two examples it is evident that
for large $t$, $y_1\sim t^0, y_2\sim t^1, y_3\sim t^2, y_4\sim e^t,
y_5\sim e^t$.
Because the populations of the 2-cycle are growing exponentially they will
eventually completely overshadow the populations of the chain which are growing
only as powers of $t$. Therefore the attractor will be ${\bf X}=(0,0,0,1,1)/2$
which, it can be checked, is a PFE of the graph (it is an eigenvector with
eigenvalue 1).

In general when a graph consists of one or more ACSs and other nodes that
are not part of any ACS,
the populations of the ACS nodes grow exponentially while the populations of
the latter nodes grow at best as powers of $t$. Hence ACSs always
outperform non-ACS structures in the population dynamics (see also 
Example 2). This is a consequence of
the infinite walks provided by the positive feedback inherent in the ACS structure,
while non-ACS structures have no feedbacks and only finite walks.\\

\noindent
{\bf Example 6}. A 2-cycle and another irreducible graph disconnected from it, Figure \ref{uniquePFEfig}f:\\
One can ask, when there is more than one ACS in the graph, which is the dominant ACS?
Figure \ref{uniquePFEfig}f shows a graph containing two ACSs. The 2-cycle subgraph has a
Perron-Frobenius eigenvalue 1, while the other irreducible subgraph has a Perron-Frobenius
eigenvalue $\sqrt{2}$. The unique PFE of the entire graph is ${\bf e} = 
(0,0,1,\sqrt{2},1)/(2+\sqrt{2})$ with eigenvalue $\sqrt{2}$.
The population dynamics equations are $\dot{y_1}=y_2, \dot{y_2}=y_1,
\dot{y_3}=y_4, \dot{y_4}=y_3+y_5, \dot{y_5}=y_4$.
The first two equations are completely decoupled from the last three and 
the solutions for $y_1$ and $y_2$ are the same as for Example 3.
For the other irreducible graph the solution is (since $\ddot{y_4}=\dot{y_3} + \dot{y_5}
= 2y_4$)
$$y_4(t)=Ae^{\sqrt{2}t} + Be^{-\sqrt{2}t}, \quad \quad
y_3(t)={1 \over \sqrt{2}}(Ae^{\sqrt{2}t} + Be^{-\sqrt{2}t}) + C,$$
$$y_5(t)={1 \over \sqrt{2}}(Ae^{\sqrt{2}t} + Be^{-\sqrt{2}t}) - C.$$ 
Thus, the populations of nodes 3,4 and 5 also grow exponentially but at
a faster rate, reflecting the higher Perron-Frobenius eigenvalue of the
subgraph comprising those nodes.
Therefore this structure eventually overshadows the 2-cycle, and the
attractor is  ${\bf X}={\bf e}$. The dominant ACS in this
case is the irreducible subgraph formed by nodes 3,4 and 5.

More generally, when a graph consists of several disconnected ACSs with
different individual $\lambda_1$, only the ACSs whose $\lambda_1$ is the 
largest (and equal to $\lambda_1(C)$) end up with non-zero relative populations
in the attractor. \\

\noindent
{\bf Example 7}. A 2-cycle downstream from another 2-cycle, Figure \ref{uniquePFEfig}g:\\
What happens when the graph contains two ACSs whose individual $\lambda_1$
equals $\lambda_1(C)$, and one of those ACSs is downstream of another?
In Figure \ref{uniquePFEfig}g nodes 3 and 4 form a 2-cycle which is downstream from another 2-cycle
comprising nodes 1 and 2. The unique PFE of this graph, with $\lambda_1=1$, is
${\bf e}=(0,0,1,1)/2$. 
The population dynamics equations are $\dot{y_1}=y_2,
\dot{y_2}=y_1, \dot{y_3}=y_4+y_2, \dot{y_4}=y_3$. 
Their general solution is:
$$y_1(t) =  Ae^t + Be^{-t}, \quad \quad y_2(t)= Ae^t - Be^{-t},$$
$$y_3(t) = {t \over 2}(Ae^t - Be^{-t})  + Ce^t + De^{-t},$$
$$y_4(t) = {t \over 2}(Ae^t + Be^{-t})  + (C-{A \over 2})e^t + ({B \over 2}-D)e^{-t}. $$
It is clear that for large $t$, $y_1 \sim e^t, y_2 \sim e^t, y_3 \sim te^t, 
y_4 \sim te^t$. While all four grow exponentially with the same rate $\lambda_1$,
as $t\rightarrow \infty$ $y_3$ and $y_4$ will overshadow $y_1$ and $y_2$.
The attractor will be therefore be ${\bf X}=(0,0,1,1)/2 = {\bf e}$.
Here the dominant ACS is the 2-cycle of nodes 3 and 4.
This result generalizes
to other kinds of ACSs: if one irreducible subgraph is downstream of another 
with the same Perron-Frobenius eigenvalue, the latter will have zero relative population
in the attractor. 

The above examples displayed graphs with a unique PFE, and illustrated 
Proposition 5 (iv). The stability of the global attractor follows from the
fact that the constants $A,B,C,D$, etc., in the above examples, which can
be traded for the initial conditions of the populations, appear nowhere in the
attractor configuration ${\bf X}$. Now we consider examples where the PFE is not unique. \\

\noindent
{\bf Example 8}. Graph with $\lambda_1=0$ and three disconnected components, Figure \ref{nonuniquePFEfig}a:\\
As mentioned in section 2 this graph has three independent PFEs, displayed in Figure \ref{nonuniquePFEfig}a.
The attractor is ${\bf X}={\bf e}_3$. This is an immediate generalization
of Example 1 above. Using the same argument as for Example 1, 
we can see that $y_i\sim t^k$ if the longest path ending at node $i$ is of
length $k$. Therefore the attractor will have nonzero components only for
nodes at the ends of the longest paths. Thus the populations of nodes 1,2,3 and 5
are constant, those of 4 and 6 increase $\sim t$ for large $t$, and of 7 as $\sim t^2$,
explaining the result. \\

\noindent
{\bf Example 9}. Several connected components containing 2-cycles, Figure \ref{nonuniquePFEfig}b:\\
Here again there are three PFEs, one for each connected component. 
The population of nodes in 2-cycles which are not downstream of other 2-cycles
(nodes 1,2,3,4,7 and 8) will grow as $e^t$. As in Example 7, Figure \ref{uniquePFEfig}g, the nodes
of 2-cycles which are downstream of one 2-cycle (nodes 5,6,9 and 10) will
grow as $te^t$. It can be verified that the populations of nodes in 2-cycles
downstream from two other 2-cycles (nodes 11 and 12) will grow as $t^2e^t$.
The pattern is clear: in the attractor only the 2-cycles at the ends of the
longest chains of 2-cycles will have non-zero relative populations,
explaining the result.\\

\noindent {\bf Example 10.} Figure \ref{digraphfig}a:\\
From previous examples it is evident how the populations will change with time for
Figure \ref{digraphfig}a. Here we list the result:
$$y_8 \sim t^0, \quad \quad y_9 \sim t^1, \quad \quad y_{10} \sim t^2,$$
$$y_1,y_2,y_3,y_4,y_5,y_{11},y_{12},y_{13},y_{20} \sim e^t,$$
$$y_6,y_7,y_{14},y_{15},y_{16},y_{17},y_{18},y_{19} \sim te^t.$$
Thus, starting from a generic initial population, only the eight nodes, 6,7,14-19, will
be populated in the attractor. This explains the comments just after the statment
of Proposition 5.

Note the structure of the dominant ACS in the above examples when $\lambda_1 > 0$. 
If there is a unique PFE in the graph, the dominant ACS is the subgraph of the 
PFE. If there are several PFEs only a subset of those may be counted as illustrated
in Examples 9 and 10, Figures \ref{nonuniquePFEfig}b and \ref{digraphfig}a, 
respectively. A general construction of
the dominant ACS for an arbitrary graph will be described elsewhere. \\

\noindent
{\bf How long does it take to reach the attractor?}\\
The timescale over which the system reaches its attractor depends on the structure
of the graph $C$. For instance in Example 2, the attractor is approached as the
population of node 1, $y_1$, overwhelms the population $y_2$. Since $y_1$ grows
exponentially
as $e^t$, the attractor is reached on a timescale $\lambda_1^{-1}=1$.
(In general, when we say that ``the timescale for the system to reach the
attractor is $\tau$", we mean that for $t>>\tau$, ${\bf x}(t)$ is
``exponentially close" to its final destination ${\bf X}\equiv\lim_{t\rightarrow\infty} {\bf x}(t)$, i.e. for all $i$, $|x_i(t)-X_i|\sim e^{-t/\tau}t^{\alpha}$, 
with some finite $\alpha$.)
In contrast, in Example 1,
the attractor is approached as $y_3$ overwhelms $y_1$ and $y_2$. Because in this case all
the populations are growing as powers of $t$, the timescale for reaching the attractor is
infinite.
When the populations of different nodes are growing at different rates,
this timescale depends on the difference in growth rate
between the fastest growing population and the next fastest growing population. 

For graphs which have no basic subgraphs, i.e., graphs with $\lambda_1=0$ like those in Example 1
and 8, all populations grow as powers of $t$, hence the timescale for reaching the attractor
is infinite.

For graphs which have one or more basic subgraphs
(i.e., $\lambda_1\ge 1$) but all the basic subgraphs are in different
connected components, such as Examples 2-6, the timescale for reaching the attractor is given by
$\left(\lambda_1- {\rm Re} \lambda_2\right)^{-1}$, where
$\lambda_2$ is the eigenvalue of $C$ with the next largest real part, compared to $\lambda_1$.

For graphs having one or more basic subgraphs with at least one basic subgraph downstream from
another basic subgraph, the ratio of the fastest growing population to the next fastest growing 
will always be a power of $t$ (as in Examples 7, 9 and 10) therefore the timescale for reaching
the attractor is again infinite.\\

\noindent
{\bf Core and periphery of a graph}\\
Since the dominant ACS is given by a PFE, 
we will define the core of the dominant ACS to be the core of the corresponding PFE.
If the PFE is simple, the core of the dominant ACS consists of just one basic subgraph.
If the PFE is non-simple the core of the dominant ACS will be a union of some basic
subgraphs.
Further, the dominant ACS is uniquely determined by the graph. This motivates
the definition of the core and periphery of a graph:
The {\it core} of a graph $C$, denoted $Q(C)$, is the 
core of the dominant ACS of $C$. The {\it periphery} of $C$ is the 
periphery of the dominant ACS of $C$. This definition applies when $\lambda_1(C) > 0$.
When $\lambda_1(C) = 0$, the graph has no ACS and by definition $Q(C)=\Phi$.
In all cases $\lambda_1(Q(C))=\lambda_1(C)$.
For all the graphs depicted in this paper, except the one in Figure \ref{digraphfig}e,
the red nodes constitute the core of the graph, the
blue nodes its periphery, and the white nodes are neither core nor periphery --
they are nodes that are not in any of the PFE subgraphs.
\footnote{The definition of the core of a graph given in refs. \cite{JK4,JK5}
is a special case of this definition, holding only for graphs where each connected
component of the dominant 
ACS has no more than one basic subgraph.}\\

\noindent
{\bf Core overlap of two graphs}\\
Given any two graphs $C$ and $C'$ whose nodes are labeled,
the {\it core overlap} between them, denoted $Ov(C,C')$, is
the number of common links in the cores of $C$ and $C'$, i.e.,
the number of ordered pairs $(j,i)$ for which $Q_{ij}$ and $Q'_{ij}$
are both non-zero \cite{JK4}. If either of $C$ or $C'$ does not have a core, $Ov(C,C')$ is 
identically zero.\\

\noindent
{\bf Keystone nodes}\\
In ecology certain species are referred to as keystone species -- those
whose extinction or removal would seriously disturb the balance of the
ecosystem \cite{Paine,Pimm,JTM,SMo}. One might similarly ask 
for the notion of a keystone node in a directed graph that captures some important
organizational role played by a node. 
Consider the impact of the hypothetical removal of any node $i$
from a graph $C$.
One can, for example, ask for the core of the graph $C-i$ that would result
if node $i$ (along with all its links) were removed from $C$.
We will refer to a node $i$ as a {\it keystone node} if $C$ has a non-vanishing
core and $Ov(C,C-i)=0$ \cite{JK5}. Thus a keystone node is one whose removal modifies the
organizational structure of the graph (as represented by its core) drastically.
In each of Figures \ref{intpathfig}a-d, for example, the core is the entire graph. 
In Figure \ref{intpathfig}a, all the nodes are keystone, since the removal of any one of them
would leave the graph without an ACS (and hence without a core). In general when the
core of a graph is a single $n$-cycle, for any $n$, all the core nodes are keystone.
In Figure \ref{intpathfig}b, nodes 3, 4 and 5 are keystone
but the other nodes are not, and in Figure \ref{intpathfig}c only nodes 4 and 5 are keystone. 
In Figure \ref{intpathfig}d, there are no keystone nodes. These examples show that the more
internal pathways a core has (generally, this implies a higher value of 
$\lambda_1$), the less likely it is to have keystone species, and hence the
more robust its structure is to removal of nodes.

Figure \ref{keystonefig} illustrates another type of graph structure which has a keystone node.
The graph in Figure \ref{keystonefig}a consists of a 2-cycle (nodes 4 and 5) downstream from 
an irreducible subgraph consisting of nodes 1,2 and 3.
The core of this graph is the latter irreducible subgraph.
Figure \ref{keystonefig}b shows the graph that results if node 3 is removed with all its
links. This consists of one 2-cycle downstream from another. Though both
2-cycles are basic subgraphs of the graph, as discussed in Example 7, Figure \ref{uniquePFEfig}g, 
this graph has
a unique (upto constant multiples) PFE, whose subgraph consists of the
downstream cycle (nodes 4 and 5) only. Thus the 2-cycle 4-5 is the core
of the graph in Figure \ref{keystonefig}b. Clearly $Ov(C,C-3)=0$ therefore node 3 in Figure \ref{keystonefig}a is
a keystone node.

We remark that the above purely graph theoretic definition of a keystone node 
turns out to 
be useful in the dynamical system discussed in this and the following sections. For other
dynamical systems, other definitions of keystone might be more useful.

\section{Graph dynamics}

So far we have discussed the algebraic properties of a fixed graph,
and the attractors of a particular dynamical system on arbitrary, but fixed graphs.
However one of the most interesting properties of complex systems is that
the graph of interactions among their components evolves with time, resulting
in many interesting adaptive phenomena. We now turn to such an example,
where the graph itself is a dynamical variable, and display how phenomena
such as self-organization, catastrophes, innovation, etc, can arise. We shall see
that the above discussion of (static) graph theory will be crucial in
understanding these phenomena.

We consider a process which alters a graph in discrete steps.
The series of graphs produced by such a process can be denoted
$C_n, n=1,2,\ldots$. A graph update event will be one step of the process, taking a graph
from $C_{n-1}$ to $C_n$. In fact the process we consider is a specific example
of a Markov process on the space of graphs. At time $n-1$, the graph
$C_{n-1}$ determines the transition probability to all other graphs. The
stochastic process picks the new graph $C_n$ using this probability distribution
and the trajectory moves forward in graph space. In the example we consider,
the transition probability is not specified explicitly. It arises implicitly as
a consequence of the dynamics (\ref{xdot}) that takes place on a fast
time scale for the fixed graph $C_{n-1}$.\\

The graph dynamics is implemented as follows \cite{JK1}:\\
Initially the graph is random: for every ordered pair $(i,j)$ with $i\ne j$,
$c_{ij}$ is independently
chosen to be unity with a probability $p$ and
zero with a probability $1-p$.
$c_{ii}$ is set to zero for all $i$.
Each $x_i$ is chosen randomly in $[0,1]$ and all $x_i$ are rescaled
so that $\sum_{i=1}^{s}x_i=1$.\\

Step 1. With $C$ fixed, ${\bf x}$ is evolved according to (\ref{xdot})
until it converges to a fixed point, denoted ${\bf X}$.
The set $\cal{L}$ of nodes with the least
$X_i$ is determined,
i.e, ${\cal{L}}=\{i \in S|X_i = {\rm min}_{j \in S} X_j \}$.\\

Step 2. A node, say node $k$, is picked randomly from $\cal{L}$ and is removed from
the graph along with all its links.\\

Step 3. A new node (also denoted $k$) is added to the graph. Links to and from $k$ to other nodes
are assigned randomly according to the same rule, i.e,
for every $i\ne k$
$c_{ik}$  and $c_{ki}$ are independently
reassigned to unity with probability $p$ and zero with probability $1-p$, irrespective
of their earlier values, and
$c_{kk}$ is set to zero. All other matrix elements of $C$ remain unchanged.
$x_k$ is set to a small constant $x_0$, all other $x_i$ are perturbed by a small amount
from their existing value $X_i$, and all $x_i$ are rescaled so that $\sum_{i=1}^s x_i=1$.\\

This process, from step 1 onwards, is iterated many times.

Notice that the population dynamics and the graph dynamics are coupled:
the evolution of the $x_i$ depends on the graph $C$ in step 1, and the evolution
of $C$ in turn depends on the $x_i$ through the choice of which node to remove
in step 2. There are two timescales in the dynamics, a short timescale over which
the graph is fixed while the $x_i$ evolve, and a longer timescale over which
the graph is changed.

This dynamics is motivated by the origin of life problem, in particular the
puzzle of how a complex chemical organization might have emerged from an initial
`random soup' of chemicals, as discussed in section 1.
Let us consider a pond on the prebiotic earth containing $s$ molecular
species which interact catalytically as discussed in the previous section,
and let us allow the chemical organization to evolve with time due to various
natural process which remove species from the pond and bring new species into
the pond.
Thus over short timescales we let
the populations of the species evolve
according to (\ref{xdot}). Over longer timescales
we imagine the prebiotic pond to be subject to periodic perturbations from
storms, tides or floods. These perturbations remove existing species from
the pond and introduce new species into it. The species most likely to
be completely removed from the pond are those that have the least number of
molecules.
The new species could have entirely different catalytic properties from
those removed or those existing in the pond.
The above rules make the idealization that the perturbation eliminates 
exactly one existing species (that has the least relative population)
and brings in one new species. The behaviour of the system does not depend crucially
on this assumption \cite{JK5}.

While in previous sections we have considered graphs with 1-cycles, the 
requirement $c_{ii}=0$ in the present section forbids 1-cycles in the graph. 
The motivation is the following: 1-cycles represent self-replicating species
(see previous section, Example 2). Such species, e.g., RNA molecules, are 
difficult to produce and maintain in a prebiotic scenario and it is generally
believed that it requires a complex self supporting molecular organization
to be in place {\it before} an RNA world, for example, can take off \cite{JSMO,Joyce}.
Thus, we wish to address the question: can we get complex molecular organizations
without putting in self-replicating species by hand in the model? As we shall see
below, this does indeed happen, since even though self-replicating individual species
are disallowed, collectively self-replicating autocatalytic sets can still
arise by chance on a certain time scale, and when they do, they trigger a wave
of self-organization in the system.

The rules for changing the graph implement {\it selection} and {\it novelty},
two important features of natural evolution. Selection is implemented by
removing the species which is `performing the worst', with `performance' in this
case being equated to a species' relative population (step 2). Adding a new species
introduces novelty into the system. Note that although the actual
connections of a new node with other nodes are created randomly,
the new node has the same average
connectivity as the initial set of nodes. Thus the new species is not biased
in any way towards increasing the complexity of the chemical organization.
Step 2 and step 3 represent the interaction of the system with the external
environment. The third feature of the model is dynamics of the system that depends
upon the interaction among its components (step 1). The phenomena to be described
in the following sections are all consequences of the interplay between these three 
elements -- selection, novelty and an internal dynamics.

\section{Self Organization}

We now discuss the results of graph evolution.
Figure \ref{linksfig} shows the total number of links in the graph versus time ($n$, the
number of graph updates). Three runs of the model described in the previous section,
each with $s=100$ and different values
of $p$ are exhibited. Also exhibited is a run where there was {\it no selection} 
(in which step 2 is modified: instead of picking one of the nodes of $\cal{L}$,
any one of the $s$ nodes is picked randomly and removed from the graph along with all
its links. The rest of the procedure remains the same).
Figure \ref{s1l1fig} shows the time evolution of two more quantities for the same three
runs with selection displayed in Figure \ref{linksfig}. The quantities plotted are the
number of nodes with $X_i>0$, $s_1$, and the Perron-Frobenius eigenvalue
of the graph, $\lambda_1$. The values of the parameters $p$ and $s$ for the 
displayed runs were chosen to lie in the regime $ps<1$. Much of the analytical
work described below, such as estimation of various timescales, assumes that
$ps<<1$.
Figure \ref{snapshotsfig} shows snapshots of the graph at various times in the run shown
in Figure \ref{s1l1fig}b, which has $p=0.0025$.
It is clear that without selection each graph update replaces a randomly chosen
node with another which has on average the same connectivity. Therefore the
graph remains random like the starting graph and the number of links
fluctuates about its random graph value $\approx ps^2$.
As soon as selection is turned on
the behaviour becomes more interesting. Three regimes can be observed.
First, the `random phase' where the number of links fluctuates around $ps^2$
and $s_1$ is small.
Second, the `growth phase' where $l$ and $s_1$ show a clear rising tendency.
Finally, the `organized phase' where $l$ again hovers (with large fluctuations)
about a value much higher than the initial random graph value, and
$s_1$ fluctuates (again with large fluctuations) about its maximum value $s$.
The time spent in each phase clearly depends on $p$, and we find it also depends
on $s$. This behaviour can be understood by taking a look at the structure of
the graph in each of these phases, especially the ACS structure, and using the results of sections 2 and 3.

\subsection{The random phase}

Initially, the random graph contains no cycles, and hence no ACSs, and its Perron-Frobenius eigenvalue is
$\lambda_1=0$. We have seen in section 3 that for such a graph the
attractor  will have nonzero components for all nodes
which are at the ends of the longest paths of nodes, and zero for every other
node.
(In Figure \ref{snapshotsfig}a, there are two paths of length 4, which are the longest paths in
the graph. Both end at node 13, which is therefore the only populated
node in the attractor for this graph.)
These nodes, then, are the only nodes protected from elimination during the
graph update. However, these nodes have high relative populations {\it because
they are supported by other nodes}, while the latter (supporting) nodes do not have
high relative populations. Inevitably within a few graph updates a supporting
node will be removed from the graph. When that happens a node which presently has nonzero $X_i$
will no longer be at the end of the longest path and hence will get $X_i=0$. 
For example node 34, which belongs to
$\cal{L}$, is expected to be picked for replacement within $\approx O(s)$
graph update time steps. In fact it is replaced in the 8th time step. After
that node 13 becomes a singleton and joins the set $\cal{L}$.
Thus no structure is stable when there is no ACS. Eventually, all
nodes are removed and replaced, and the graph remains random.

Note that the inital random graph is likely to contain no cycles when $p$ is small
($ps<<1$). If larger values of $p$ are chosen, it becomes more likely that the 
initial graph will contain a cycle. If it does, there is no random phase; the
system is then in the growth phase, discussed below, right from the initial time step.

\subsection{The growth phase}

At some graph update an ACS is formed by pure chance. The probability of this
happening can be closely approximated by the probability of a 2-cycle (the
simplest ACS) forming by chance, which is $p^2s$ (= the probability that in the
row and column corresponding to the replaced node in $C$, any matrix element
and its transpose both turn out to be unity). Thus the average time of
appearance of an ACS is $1/p^2s$. In the run whose snapshots are displayed
in Figure \ref{snapshotsfig}, a 2-cycle between nodes
26 and 90 formed at $n=2854$. This is a graph which consists of a 2-cycle and
several other chains and trees. For such a graph we have shown in Example 3 in
section 3 that the attractor has non-zero $X_i$
for nodes 26 and 90 and zero for all other nodes. The dominant ACS consists
of nodes 26 and 90. Therefore these nodes cannot be picked for removal at the
graph update and hence a graph update cannot destroy the links that make
the dominant ACS. {\it The autocatalytic property is guaranteed to be preserved
until the dominant ACS spans the whole graph}.

When a new node is added to the graph at a graph update, one of three things
will happen:

1. The new node will not have any links from the dominant ACS and will not
form a new ACS. In this case the dominant ACS will remain unchanged, the new
node will have zero relative population and will be part of the least fit set.
For small $p$ this is the most likely possibility.

2. The new node gets an incoming link from the dominant ACS and hence becomes a
part of it. In this case the dominant ACS grows to include the new node.
For small $p$, this is less likely than the first possibility,
but such events do happen and in fact are the ones responsible for the
growth of complexity and structure in the graph.

3. The new node forms another ACS. This new ACS competes with the existing
dominant ACS. Whether it now becomes dominant, overshadowing the previous
dominant ACS or it gets overshadowed, or both ACSs coexist depends on the
Perron Frobenius eigenvalues of their respective subgraphs and whether (and which)
ACS is downstream of the other. It can be shown that this is a rare event compared
with possibilities 1 and 2.

Typically the dominant ACS keeps growing by accreting new nodes, usually one at a time,
until the entire graph is an ACS. At this point the growth phase stops and the
organized phase begins. As a consequence it follows that {\it $\lambda_1$ is a 
nondecreasing function of $n$ as long as $s_1<s$} \cite{JK2}.\\

\noindent
{\bf Time scale for growth of the dominant ACS.}\\
If we assume that possibility 3 above is rare enough to neglect, and that
the dominant ACS grows by adding a single node at a time, we can estimate
the time required for it to span the entire graph. Let the dominant ACS
consist of $s_1(n)$ nodes at time $n$. The probability that the new node
gets an incoming link from the dominant ACS and hence joins it is $ps_1$.
Thus in $\Delta n$ graph updates, the dominant ACS will grow, on average,
by $\Delta s_1=ps_1\Delta n$ nodes. Therefore
$s_1(n)=s_1(n_a)exp((n-n_a)/\tau_g)$,
where $\tau_g=1/p$, $n_a$ is the time of arrival of the first ACS and
$s_1(n_a)$ is the size of the first ACS (=2 for the run shown in Figure \ref{snapshotsfig}). 
Thus $s_1$ is expected to grow
exponentially with a characteristic timescale $\tau_g=1/p$. The time taken
from the arrival of the ACS to its spanning is $\tau_g \ln (s/s_1(n_a))$.
This analytical result is confirmed
by simulations (see Figure \ref{taugfig}).

In the displayed run, after the first ACS (a 2-cycle) is formed at $n=2854$,
it takes 1026 time steps, until $n=3880$ for the dominant ACS to span the entire
graph (Figure \ref{snapshotsfig}c).
This explains how an autocatalytic network structure and the positive feedback
processes inherent in it can bootstrap themselves into existence from a small
seed. The small seed, in turn, is more or less guaranteed to appear on a certain
time scale ($1/p^2s$ in the present model) just by random processes.
\\

\noindent
{\bf A measure of the `structure' of the evolved graph.}\\
A fully autocatalytic graph is a highly improbable structure.
Consider a graph of $s$ nodes and let the probability of a positive link
existing between any pair of nodes be $p^*$. Such a graph has on average
$m^*=p^*(s-1)$ incoming or outgoing positive links per node since links from
a node to itself are disallowed.
For the entire graph to be an ACS, each node must have at least one
incoming link, i.e. each row of the matrix $C$ must contain at least one
positive element. Hence the probability, $P$, for the entire graph to be an
ACS is\\
\begin{tabular}{ccl}
$P$ & $=$ & probability that every row has at least one positive entry\\
    & $=$ & [probability that a row has at least one positive entry$]^s$\\
    & $=$ & $[1-($probability that every entry of a row is zero$)]^s$\\
    & $=$ & $[1-(1-p^*)^{s-1}]^s$\\
    & $=$ & $[1-(1-m^*/(s-1))^{s-1}]^s$\\
\end{tabular}

Note from Figure \ref{linksfig} that at spanning the number of links is $O(s)$. Thus the average
degree $m^*$ at spanning is $O(1)$. We have found this to be true in all the
runs we have done where the initial average degree (at $n=1$)
was $O(1)$ or less.

For large $s$ and $m^*\sim O(1)$,
$P\approx (1-e^{-m^*})^s\sim e^{-\alpha s}$,
where $\alpha$ is positive, and $O(1)$.
Thus a fully autocatalytic graph is exponentially unlikely to form if it
were being assembled randomly. In the present model nodes are being
added completely randomly but the underlying population dynamics and the
selection imposed at each graph update result in the inevitable arrival
of an ACS (in, on average, $\tau_a=1/p^2s$ time steps) and its
inevitable growth into a fully autocatalytic graph in (on average) an
additional $\sim\tau_g \ln s$ time steps.

It is a noteworthy feature of self-organization in the present model that an
organization whose a priori probability to arise is exponentially small,
$\sim e^{-\alpha s}$, arises inevitably in a rather short time,
$\sim \frac{1}{p}\ln s$ (for large $s$).
Why does that happen? First a small ACS
of size $s_1(n_a)\sim O(1)$ forms by pure chance. The probability of this
happening is not exponentially small; it is in fact quite substantial.
Once this has formed, it is a cooperative structure and is
therefore stable. Its appearance ushers in an exponential growth of structure
with a time scale $\tau_g=1/p$. Hence a graph whose `structuredness'
(measured by the reciprocal
of the probability of its arising by pure chance) $=e^{\alpha s}$ arises in only $\frac{1}{p} \ln s$ steps.

As mentioned in the introduction, one of the major puzzles in the origin
of life is the emergence of 
very special chemical organizations in a relatively short time. 
We hope that the mechanism described above, or its analogue
in a sufficiently realistic model, will help in addressing this puzzle.
The relevance of this mechanism for the origin of life
is discussed in ref. \cite{JK3}. 
We remark that other models of self-organization
(e.g. the well-stirred hypercycle) do not seem to be able to produce
complex structured organizations from a simple starting network (see ref. \cite{JK5}). 

Another graph theoretic measure of the structure of the evolved graph is 
`interdependency' among the nodes, discussed in \cite{JK2,JK3}.
Like the links and $s_1$, the interdependency is low in the random phase,
then rises in the growth phase to a value that is about an order of magnitude
higher.

\subsection{The organized phase}

Once an ACS spans the entire graph the effective dynamics again changes
although the microscopic dynamical rules are unchanged.
At spanning, for the first time since the
formation of the initial ACS, a member of the dominant ACS will be picked for removal.
This is because at spanning all nodes by definition belong to the dominant ACS
and have non zero relative populations;
one node nevertheless has to be picked for removal.
Most of the time the removal of the node with the least $X_i$ will result 
in minimal damage to the ACS. The rest of the ACS will remain with high 
populations, and the new node will keep getting repeatedly removed and replaced
until it once again joins the ACS. Thus $s_1$ will fluctuate between $s$ and
$s-1$ most of the time. However, once in a while, the node which is removed
happens to be playing a crucial role in the graph structure despite its low
population. Then its removal can trigger large changes in the structure and
catastrophic drops in $s_1$ and $l$. Alternatively it can sometimes happen
that the new node added can 
trigger a catastrophe because of the new graph structure it creates.
The catastrophes and the mechanisms which cause them 
are the subject of the next section.

\section{Catastrophes and recoveries in the organized phase}

Figure \ref{longtimefig} shows the same run as that of Figure \ref{s1l1fig}b for $n=1$ to $n=$50,000.
In this long run one can see several sudden, large drops in $s_1$: {\it catastrophes} in which 
a large fraction of the $s$ species become extinct.
Some of the drops seem to take the system back into the random phase, others are
followed by {\it recoveries} in which $s_1$ rises back towards its maximum value $s$.
The recoveries are comparatively slower than the catastrophes, which in fact occur
in a single time step.

In order to understand what is happening during the catastrophes and subsequent recoveries
we begin by examining the possible changes that an addition or a 
deletion of a node can
make to the core of the dominant ACS.\\

\noindent
{\bf Deletion of a node}\\
We have already seen how the deletion of a node can change the core -- recall
the discussion of keystone nodes in section 3:
the removal of a keystone node results in a zero overlap between the cores of the
dominant ACS before and after the removal.
A zero core overlap means that a single graph update event (in which one of the least
populated species is replaced by a randomly connected one) has caused a major
reorganization of the dominant ACS: the cores of the dominant ACS before and 
after the event (if an ACS still exists) have not even a single link in common. 
We will call such events {\it core-shifts}.

In an actual run a keystone node can only be removed if it happens to be one
of the nodes with the least $X_i$. However the core nodes are often `protected'
by having higher $X_i$. Why is that?

${\bf X}$ is an eigenvector of $C$
with eigenvalue $\lambda_1$. Therefore, when $\lambda_1 \neq 0$
it follows that for nodes of the dominant ACS,
$X_i=(1/{\lambda_1})\sum_j c_{ij}X_j$.
If node $i$ of the dominant ACS has only one incoming link
(from the node $j$, say)
then $X_i=X_j/\lambda_1$; we can say that $X_i$ is `attenuated' with respect to
$X_j$ by a factor $\lambda_1$.                                                  
The periphery of an ACS is a tree like structure emanating
from the core, and for small $p$ most periphery nodes have a single incoming link.
For instance the graph in Figure \ref{snapshotsfig}c, whose $\lambda_1=1.31$,
has a chain of nodes $44\rightarrow 45\rightarrow 24\rightarrow 29\rightarrow 52\rightarrow
89\rightarrow 86\rightarrow 54\rightarrow 78$.
The farther down such a chain a periphery node is, the lower is its
$X_i$ because of the cumulative attenuation. For such an ACS with $\lambda_1>1$
the `leaves' of the periphery tree (such as node 78) 
will typically be the species with
least $X_i$ while the core nodes will have larger $X_i$.

However, when $\lambda_1=1$ there is no attenuation. 
Recall that Proposition 1(iii) shows that at $\lambda_1=1$ the core must be a cycle
or a set of disjoint cycles,
hence each core node has only one incoming link within the dominant ACS.
All core nodes have the same value of $X_i$. As one moves out
towards the periphery $\lambda_1=1$ implies there is no attenuation, hence each node
in the periphery that receives a single link from one of the core nodes will also
have the same $X_i$.
Some periphery nodes may have higher $X_i$ if they have more than
one incoming link from the core. Iterating this argument as one moves further outwards from
the core, it is clear that at $\lambda_1=1$ the core is not protected and in fact
will always belong to the set of least fit nodes 
if the dominant ACS spans the graph. We have already seen in section 3 that when 
$\lambda_1=1$ and the core is a single cycle every core node is a keystone node. Thus when $\lambda_1=1$ the
organization is fragile and susceptible to core-shifts caused by the removal of
a keystone node. \\

\noindent
{\bf Addition of node}\\
We now turn to the effects of the addition of a node to the dominant ACS.
We will use the notation $C_n' \equiv C_{n-1} - k$ for
the graph of
$s-1$ nodes just before the new node at time step $n$ is brought in (and just
after the least populated species $k$ is removed from $C_{n-1}$).
$Q_n'$ will stand for the core of $C_n'$. 
In the new attractor the new species $k$ may go
extinct, i.e., $X_k$ may be zero, or it may survive, i.e., $X_k$ is non-zero.
If the new species goes extinct then it remains in the set of least fit nodes
and clearly there is no change to the dominant ACS. So we will
focus on events in which the new species survives in the new attractor.\\

\noindent
{\bf Innovations}\\
We define an {\it innovation} to be a new node for which 
$X_k$ in the new attractor is nonzero, i.e. a new node which survives
till the next graph update \cite{JK5}. This may seem to be a very weak requirement,
yet we will see that it has nontrivial consequences. 
A description of various types of innovations and their consequences, with examples, is given
in \cite{JK6}. Here we present a graph-theoretic classification of 
innovations (in terms of a hierarchy, see Figure \ref{hierarchyfig}).

The innovations which have the least impact on the populations of the
species and the evolution of the graph on a short time scale 
(of a few graph updates) are ones which do not
affect the core of the dominant ACS, if it exists. Such innovations are of three
types (see boxes 1-3 in Figure \ref{hierarchyfig}):\\
{\bf 1. Random phase innovations.} These are innovations which occur in the random phase
when no ACS exists in the graph, and they do not create any new ACSs. These innovations
are typically short lived and have little short term or long term impact on the structure
of the graph.\\
{\bf 2. Incremental innovations.} These are innovations which occur in the growth
and organized phases, which add new nodes to the periphery of the dominant ACS
without creating any new irreducible subgraph.
In the short term they only affect the periphery and are responsible for the growth
of the dominant ACS. In a longer term they can also affect the core as chains of nodes
from the periphery join the core of the dominant ACS.\\
{\bf 3. Dormant innovations.} These are innovations which occur in the growth and 
organized phases, which create new irreducible subgraphs in the periphery of the dominant 
ACS. These innovations also affect only the periphery in the short term. But they
have the potential to cause core-shifts later if the right conditions occur
(discussed in the next subsection).\\

Innovations which do immediately affect the core of the existing dominant ACS are 
always ones which create a new irreducible subgraph. They are also of three types
(see boxes 4-6 in Figure \ref{hierarchyfig}):\\
{\bf 4. Core enhancing innovations.} These innovations result in the expansion
of the existing core by the addition of new links and nodes from the periphery or 
outside the dominant ACS. They result in an increase of $\lambda_1$ of the graph.\\
{\bf 5. Core-shifting innovations.} These are innovations which cause an immediate core-shift
often accompanied by the extinction of a large number of species.\\
{\bf 6. Creation of the first ACS.} This is an innovation which creates an ACS for the first time
in a graph which till then had no ACSs. The innovation moves the system from the random phase
to the growth phase, triggering the self organization of the system around the newly created ACS.\\

Innovations of types 4, 5 and 6 which affect the core of the dominant ACS will be
called {\it core-transforming innovations}. These innovations cause a substantial change
in the vector of relative populations in a single graph update. Innovations of type 5 and 6
also make a qualitative change in the structure of the graph and significantly 
influence subsequent
graph evolution. The following theorem makes precise the conditions under which a core
transforming innovation can occur.\\

\noindent
{\bf Core transforming Theorem}\\
Let $N$ (or $N_n$ at time step $n$)
denote the maximal new irreducible
subgraph which includes the new species.  
One can show that $N_n$ will become the new
core of the graph, replacing the old core $Q_{n-1}$, whenever
either of the following conditions are true: \\
(a) $\lambda_1(N_n)>\lambda_1(Q_n')$ or, \\
(b)
$\lambda_1(N_n)=\lambda_1(Q_n')$ and $N_n$ is `downstream'
of $Q_n'$ (i.e., there is a path from $Q_n'$ to $N_n$ but not
from $N_n$ to $Q_n'$.)

Such an innovation will fall into category 4 above if $Q_{n-1} \subset N_n$.
However, if $Q_{n-1}$ and $N_n$ are disjoint, we get a core-shift and the
innovation is of type 5 if $Q_{n-1}$ is non-empty and type 6 otherwise.\\

\subsection{Catastrophes, core-shifts and a classification of proximate causes}
The large sudden drops visible in Figure \ref{longtimefig} are now discussed. Our first task is 
to see if the large drops are correlated to specific changes in the structure
of the graph.
Let us focus on those events in which more than 50\% of the species go extinct.
There were 701 such events out of 1.55 million graph updates 
in a set of runs with $s=100, p=0.0025$.
Figure \ref{histogramfig} shows a histogram of core overlaps $Ov(C_{n-1},C_n)$
for these 701 events. 612 of these have zero core overlap, i.e., they are core-shifts.
If we now look at only those events in which more than 90\% of the species
went extinct then we find 235 such events in the same runs, out of which 226
are core-shifts. Clearly most of the large extinction events happen when there
is a drastic change in the structure of the dominant ACS -- a core-shift.\\
                                                                                
\noindent
{\bf Classification of core-shifts}\\ 
Using the insights from the above discussion of the effects of deletion or addition of a node,
we can classify the different mechanisms which cause
core-shifts.
Figure \ref{classifig} differentiates between the 612 core-shifts we observed amongst the
701 crashes. 
They fall into three categories \cite{JK5}: (i) complete 
crashes
($136$ events), 
(ii) takeovers by core-transforming innovations
($241$ events),
and (iii) takeovers by dormant innovations
($235$ events).\\

\noindent
{\bf Complete crashes}\\
A {\it complete crash} is an event in which an ACS exists before but not after 
the graph update. Such an event takes the system into the random phase.
A complete crash occurs when a keystone node is removed from the graph.
For example at $n=8232$ the graph had $\lambda_1=1$ and its core was the simple
3-cycle of nodes 20, 50 and 54. As we have seen above, when the core is a single cycle
every core node is a keystone node and is also in the set of least fit nodes.
At $n=8233$ node 54 was removed thus disrupting the 3-cycle. The resulting
graph had no ACS and $\lambda_1$ dropped to zero. As we have discussed earlier,
graphs with $\lambda_1=1$ are the ones which are most susceptible to complete
crashes. This can be seen in Figure \ref{classifig}: every complete crash occurred from
a graph with $\lambda_1(C_{n-1})=1$.\\

\noindent
{\bf Takeovers by core-transforming innovations}\\ 
An example of a takeover by a core-transforming innovation is given in 
Figures \ref{snapshotsfig}g,h. At $n=6061$ the core was a single loop
comprising nodes $36$ and $74$. Node $60$ was replaced by a new species at $n=6062$ creating a cycle
comprising nodes $60, 21, 41, 19$ and $73$, downstream from 
the old core. 
The graph at $n=6062$ has one cycle feeding into a second cycle that
is downstream from it. We have already seen in section 3 (see the discussion of
Example 4) that for such a graph
only the downstream cycle is populated and the upstream cycle and all nodes 
dependent on it go extinct. Thus the new cycle 
becomes the new core and the old core goes extinct
resulting in a core-shift.
This is an example of 
condition (b) for a core-transforming innovation.
For all such events in Figure \ref{classifig}, $\lambda_1(Q_n')=\lambda_1(C_{n-1})$ since
$k$ happened not to be a core node of $C_{n-1}$. 
Thus these core-shifts satisfy
$\lambda_1(C_n)=\lambda_1(N_n)\ge\lambda_1(Q_n')=\lambda_1(C_{n-1})\ge 1$ in 
Figure \ref{classifig}.\\

\noindent
{\bf Takeovers by dormant innovations}\\
We have earlier discussed dormant innovations, which create an irreducible
structure in the periphery of the dominant ACS which does not affect its core
at that time. For example the 2-cycle comprising nodes 36 and 74 formed
at $n=4696$. At a later time such a dormant innovation can 
result in a core-shift if the old core gets sufficiently weakened.

In this case the core has become weakened by $n=5041$, when it 
has $\lambda_1=1.24$. The structure of the graph at this time is very similar
to the graph in Figure \ref{keystonefig}a. Just as node 3 in Figure \ref{keystonefig}a was a keystone node, here
nodes 44, 85, and 98 are keystone nodes because removing any of them results in
a graph like Figure \ref{keystonefig}b, consisting of two 2-cycles, one downstream from the other.
 
Indeed at $n=5041$, node $85$ is hit and the resulting graph at $n=5042$ has 
a cycle ($26$ and $90$) feeding into another cycle ($36$ 
and $74$). Thus at $n=5042$ nodes
$36$ and $74$ form the new core with only one other
downstream node, $11$, being populated. All other nodes become depopulated
resulting in a drop in $s_1$ by $97$. 
A dormant innovation can takeover as the new core
only following a keystone extinction which weakens the old core.
In such an event the new core necessarily has a lower (but nonzero)
$\lambda_1$ than the old core, i.e., $\lambda_1(C_{n-1})>\lambda_1(C_n)\ge 1$
(see Figure \ref{classifig}).

Note that $85$ is a keystone node, and the graph is susceptible to a core-shift
{\it because} of the innovation which created the cycle 36-74 earlier.
If the cycle between $36$ and $74$
were absent, $85$ would {\it not} be a keystone species by our 
definition, since its removal would still leave part of the
core intact (nodes $26$ and $90$). \\

\subsection{Recoveries}
After a complete crash the system is back in the random phase.
In $O(s)$ graph updates each node is removed and replaced by
a randomly connected node, resulting in a graph as random as the initial graph.
Then the process starts again, with a new ACS being formed 
after an average of $1/p^2s$ time steps and then growing to span the entire graph
after, on average, $(1/p)\ln (s/s_0)$ time steps, where $s_0$ is the size of the
initial ACS that forms in this round (typically $s_0=2$).

After other catastrophes, an ACS always survives. In that case the system is
in the growth phase and immediately begins to recover, with $s_1$ growing
exponentially on a timescale $1/p$. Note that these recoveries happen because
of innovations (mainly of type 2 and 4, and some of type 3).\\

\subsection{Correlation between graph theoretic nature of perturbation
and its short and long term impact}
In previous sections we have analysed several examples of perturbations to the system.
These can be broadly placed in two classes based on their effect on $s_1$:\\
(i) `Constructive perturbations': these include the birth of a new organization (an innovation of 
type 6), the attachment of a new node to the core (an innovation of type 4) and
an attachment of a new node to the periphery of the dominant ACS (an innovation of type 2).\\
(ii) `Destructive perturbations': these include complete crashes and takeovers by dormant innovations
(both caused by the loss of a keystone node), and takeovers by core-transforming
innovations (innovations of type 5). Note that the word `destructive' is used
only in the sense that several species go extinct on a short time scale 
(a single graph update in the present model) after such a perturbation.
In fact, over a longer time scale (ranging from a few to several hundred
graph updates in the run of Figure \ref{s1l1fig}b), the `destructive' 
takeovers by innovations generally trigger a new round of 
`constructive' events like incremental innovations
(type 2) and core enhancing innovations (type 4). 

Note that the maximum upheaval is caused by those perturbations that introduce new irreducible
structures in the graph
(innovations of type 4, 5 and 6) or those that destroy the existing irreducible structure. 
For example the creation of the first ACS at $n=2854$
triggered the growth phase, a complete change in the effective dynamics of the system. 
Other examples 
of large upheavals are core-shifts caused by a takeover by a core-transforming innovation at 
$n=6061$, takeover by a dormant innovation at $n=5041$, and a complete crash at $n=8233$.
In sections 2 and 3 we have mentioned that irreducibility is related to the existence of
positive feedback and cooperation, and the `magnitude' of the feedback is measured by $\lambda_1$.
While the present model is a highly simplified model of evolving networks, we expect
that this qualitative feature, namely, the correlation between the dynamical impact
of a perturbation and its `structural' character embodied in its effect on the
`level of feedback' in the underlying graph, will hold for several other complex
systems.

\section{Concluding remarks}

In this article we have attempted to show that a certain class of dynamical
systems, those in which graphs coevolve with other dynamical variables 
living on them (in our example, living on the nodes of the graph), 
possess rich dynamical behaviour which is analytically and computationally tractable. 
Even in the highly idealized model
discussed here, this behaviour is reminiscent of what happens in real
life ---- birth of organizational structure characterized by interdependence
of components, 
cooperation of parts of the organization giving way to competition,
robust organizations becoming fragile, crashes and recoveries, innovations
causing growth as well as collapse, etc.\\ 

From the point of view of the origin of life problem the main conclusions are:\\
(i) The model shows the emergence of an organization where none
exists: a small ACS emerges spontaneously by random processes and then triggers the self-organization
of the system.\\
(ii) A highly structured organization, whose timescale of forming by pure chance is 
exponentially large (as a function of the size of the system), forms in this model 
in a very short timescale that grows only
logarithmically with the size of the system.
In \cite{JK3} we have speculated that this timescale may be $\sim 100$ million years
for peptide based ACSs, which is in the same ballpark as the timescale on which 
life is believed to have originated on the prebiotic earth.

We remark that this speculation is not necessarily in conflict with, and is possibly
complementary to, some other approaches to the origin of life:\\
(i) Complex autocatalytic organizations of polypeptides could enter into symbiosis with
the autocatalytic citric acid cycle proposed in \cite{MKYC}. The latter would
help produce, among other things, amino acid monomers needed by the former; the former 
would provide catalysts for the latter.\\
(ii) It is conceivable that membranes (possibly lipid membranes, which have been
argued to have their own catalytic dynamics \cite{SBDL}) could form in regions
where autocatalytic sets of the kind discussed here existed, thereby surrounding
complex molecular organization in an enclosure.
These `cells' may have contained different parts of the ACS, thereby endowing them 
with different fitnesses. Such an assembly could evolve. \\
(iii) It is also conceivable that
such molecular organizations formed an enabling environment for self replicating
molecules such as those needed for an RNA world.\\
Testing some of these possibilities is a task for future models and experiments.
Furthermore, the mathematical ideas and mechanisms discussed here might be relevant
for these other approaches also.\\

The present model has a number of simplifying features which depart from
realism but enhance analytical tractability. One is the linearity of the
populations dynamics on a fixed graph. Equation (\ref{xdot}) is nonlinear, but
since it originates via a nonlinear change of variables from a linear 
equation, equation (\ref{ydot}), its attractors can be easily analysed in
terms of the underlying linear system. The attractors are always fixed
points, and are just the Perron Frobenius eigenvectors of the adjacency matrix
of the graph. This allows us to use (static) graph theoretic results for
analysis of the dynamics. 

In this context it is helpful to note that while the population dynamics
in the present model is essentially linear as long as the graph is fixed, 
the model feeds the result of the population dynamics into the subsequent
graph update (the least populated node is removed). Thus over long time
scales over which the graph changes, the `coupling constants' $c_{ij}$ in
equation (\ref{xdot}) are not constant but implicitly depend upon the $x_i$,
thus making the evolution highly nonlinear. 
By virtue of the simplifying device of widely
separated time scales for the graph dynamics and the population dynamics
(the population variables reach their 
attractor before the graph is modified), what we have is piecewise linear
population dynamics. It is essentially linear between two graph updates,
and nonlinear over longer time scales because of the intertwining of
population dynamics and graph dynamics. 
This nonlinearity is essential for all the complex phenomena described above,
while the short time scale linearity is an aid in analysis. 
It would be interesting to explore complex phenomena in 
models in which the short term
population dynamics is also inherently nonlinear.
This naturally arises in prebiotic chemistry when the concentration of the 
reactants (which are assumed buffered here) are dynamical variables in addition
to the catalysts and products, as well as in several other fields.

The present model describes a well-stirred reactor; there are no spatial
degrees of freedom. This precludes a discussion of the origin of spatial
structure and its consequences alluded to in section 1. It is worthwhile to 
extend the model in that direction.
Another issue is the generation of novelty. Here
the links of the new node are drawn from a fixed probability distribution.
In real systems this distribution depends upon the (history of) states
of the system. 
A further direction for generalization consists in letting the two time 
scales of the
population and graph dynamics, separated by hand in the present model,
be endogenous.\\ 

\noindent
{\bf Acknowledgements.}
S. J. acknowledges the Associateship of the Abdus Salam International Centre for 
Theoretical Physics, Trieste.
S. K. acknowledges a Junior Research Fellowship from the Council
of Scientific and Industrial Research, India. This work is supported in
part by a grant from the Department of Science and Technology, Govt. of India.\\

\pagebreak
\vspace{-1in}
\begin{vchfigure}
\epsfbox{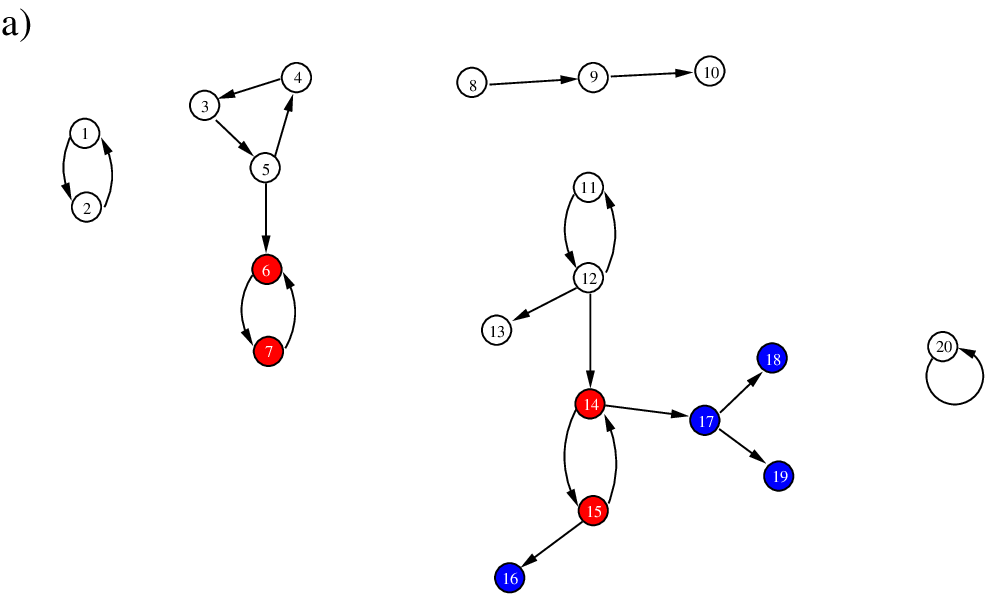}
\epsfbox{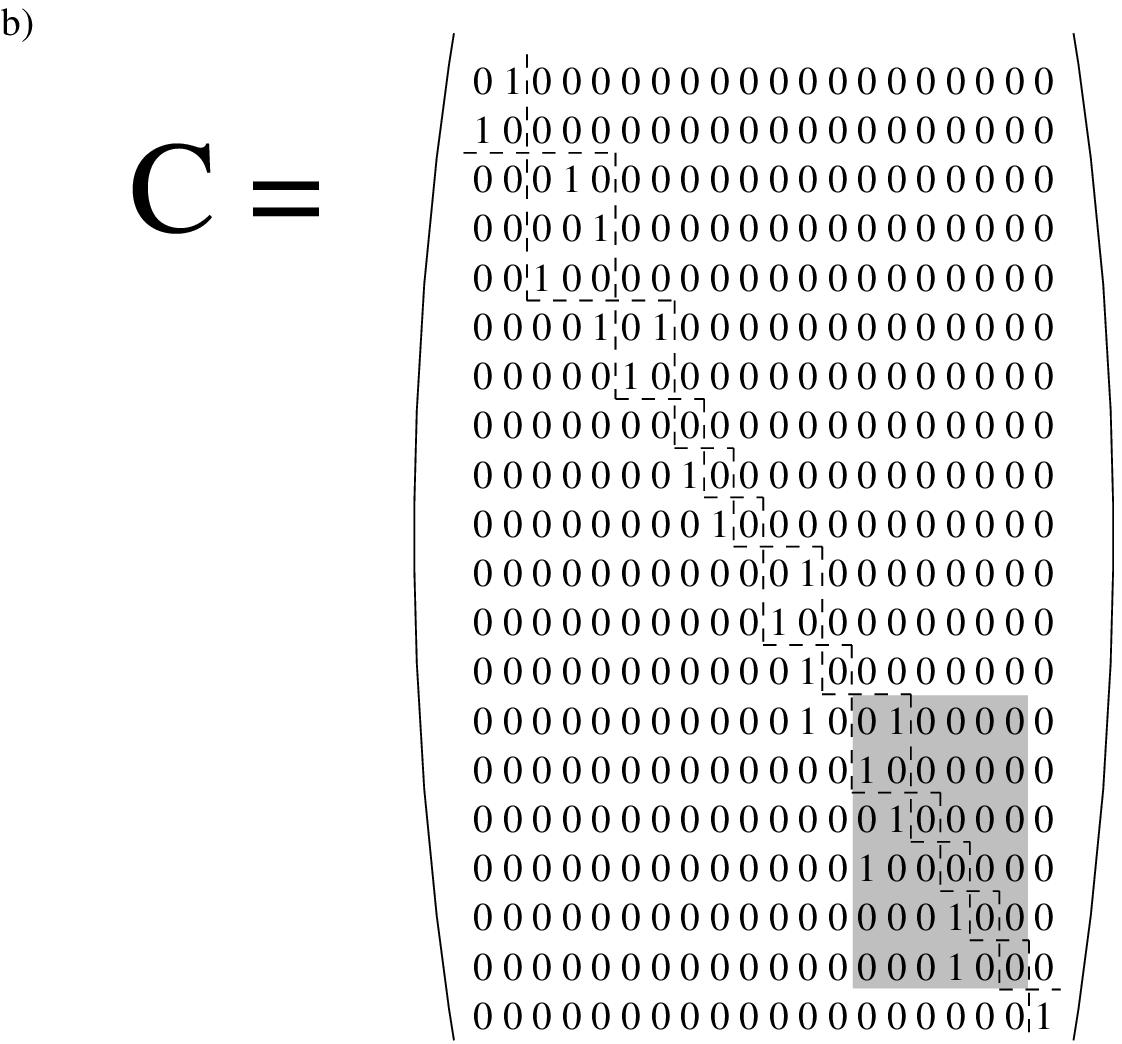}
\epsfbox{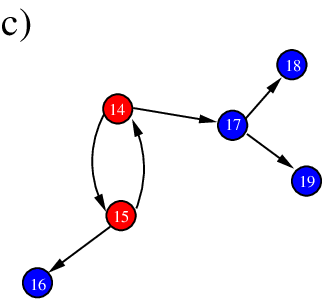}
\end{vchfigure}

\begin{vchfigure}
\epsfbox{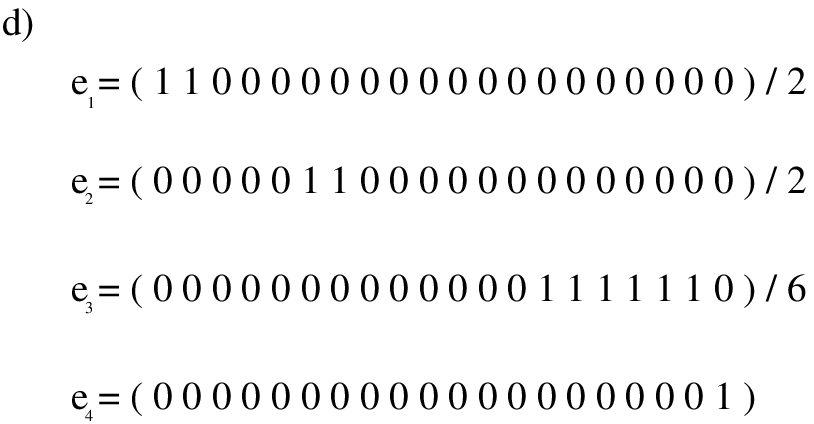}

\vspace{1in}
\epsfbox{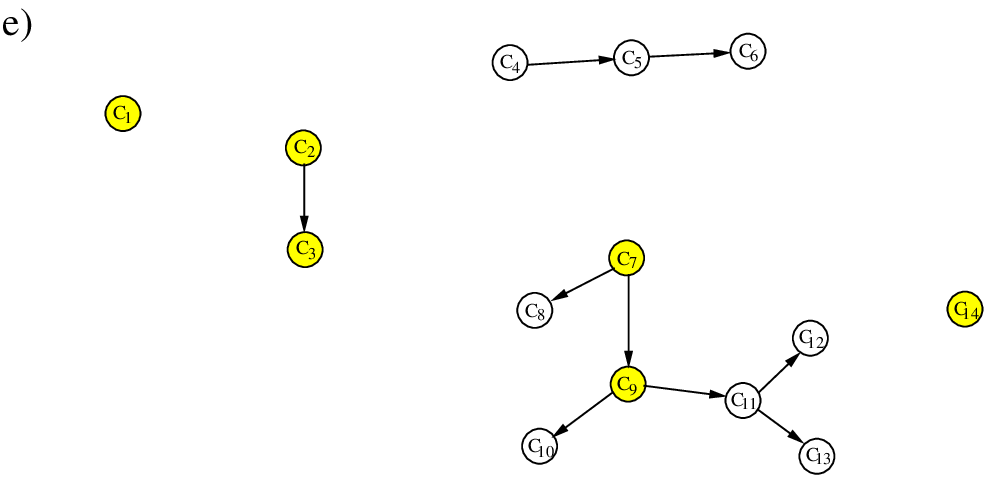}
\vchcaption{{\bf a.} A directed graph with 20 nodes.
{\bf b.} The adjacency matrix of the graph in Figure \ref{digraphfig}a. 
{\bf c.} A subgraph of the graph in Figure \ref{digraphfig}a. The adjacency matrix
of the subgraph is the shaded portion of the matrix in Figure \ref{digraphfig}b.
{\bf d.} Four Perron-Frobenius eigenvectors (PFEs) of the graph in Figure
\ref{digraphfig}a. The first three vectors have been divided by factors of 2, 2 and 6
respectively to normalize them.
{\bf e.} The irreducible decomposition of the graph in Figure \ref{digraphfig}a into subgraphs
$C_{\alpha}$, with $\alpha = 1,2, \ldots, 14$.
Each of the 14 nodes of the graph in Figure \ref{digraphfig}e represents 
either an irreducible subgraph of the graph in Figure \ref{digraphfig}a, or a single
node that is not part of any irreducible graph. The basic subgraphs of
the graph in Figure \ref{digraphfig}a are represented by yellow nodes. The dotted lines
in Figure \ref{digraphfig}b demarcate the adjacency matrices corresponding to the 
subgraphs $C_{\alpha}$.
Colours identify the attractor of the dynamics discussed in section 3,
except in Figure \ref{digraphfig}e.
In all graphs in the article (except Figure \ref{digraphfig}e), white nodes have zero
relative population in the attractor, $X_i=0$, while blue and red nodes have
$X_i > 0$. In graphs that have an autocatalytic set,  
red nodes belong to the core of the dominant autocatalytic set
of the graph, blue nodes to its periphery, and white nodes are
outside the dominant autocatalytic set.
}
\label{digraphfig}
\end{vchfigure}

\begin{vchfigure}
\epsfbox{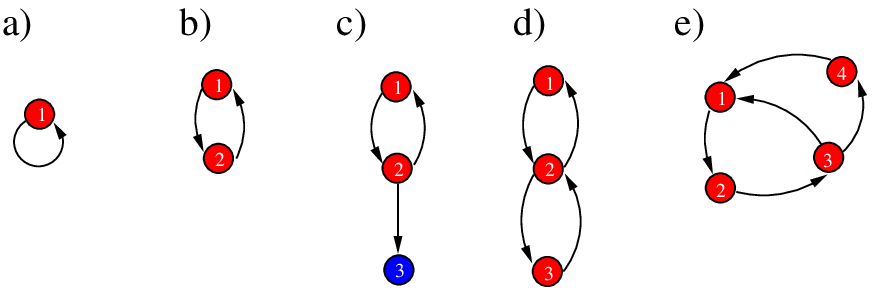}
\vchcaption{Various autocatalytic sets (ACSs). {\bf a.} A 1-cycle, the simplest
ACS. {\bf b.} A 2-cycle. {\bf c.} An ACS which is not an irreducible graph. {\bf d,e} 
Examples of ACSs which are irreducible graphs but not cycles.}
\label{acsfig}
\end{vchfigure}

\begin{vchfigure}
\epsfbox{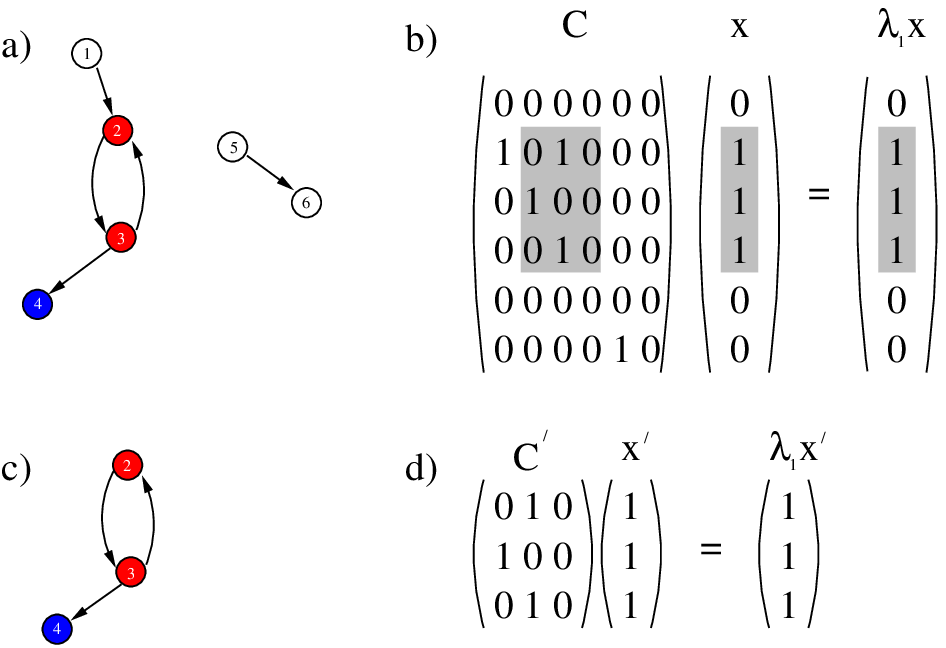}
\vchcaption{Example showing that the $\lambda_1$ of a PFE subgraph
equals the $\lambda_1$ of the whole graph.
{\bf a.} A directed graph with 6 nodes. {\bf b.} {\bf x} is an 
eigenvector of its adjacency matrix $C$ with eigenvalue $\lambda_1=1$,
which is the Perron-Frobenius eigenvalue of the graph.
The non zero components of ${\bf x}$ and the corresponding rows and columns
of $C$ are highlighted. {\bf c.} The subgraph of the PFE ${\bf x}$. 
{\bf d.} The vector {\bf x}' constructed by removing the zero components of
${\bf x}$ is an eigenvector of the adjacency matrix, $C'$, of the PFE subgraph.
Its corresponding eigenvalue is unity, which is also the Perron-Frobenius eigenvalue
of the PFE subgraph.} 
\label{PFEfig}
\end{vchfigure}

\begin{vchfigure}
\epsfbox{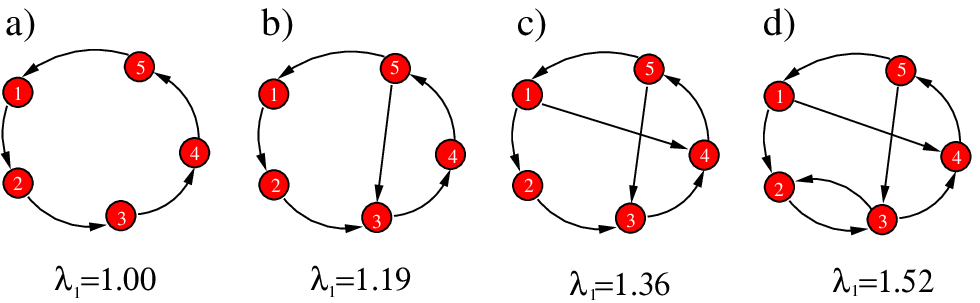}
\vchcaption{$\lambda_1$ is a measure of the multiplicity of internal pathways 
in the core of simple
PFE. Four irreducible graphs are shown. An irreducible
graph always has a unique PFE that is simple and whose core is the entire graph.
The Perron-Frobenius theorem ensures that adding a link to the core of 
a simple PFE necessarily increases its
Perron-Frobenius eigenvalue $\lambda_1$. The figure also illustrates
the concept of keystone nodes (see section 3).}
\label{intpathfig}
\end{vchfigure}

\begin{vchfigure}
\epsfbox{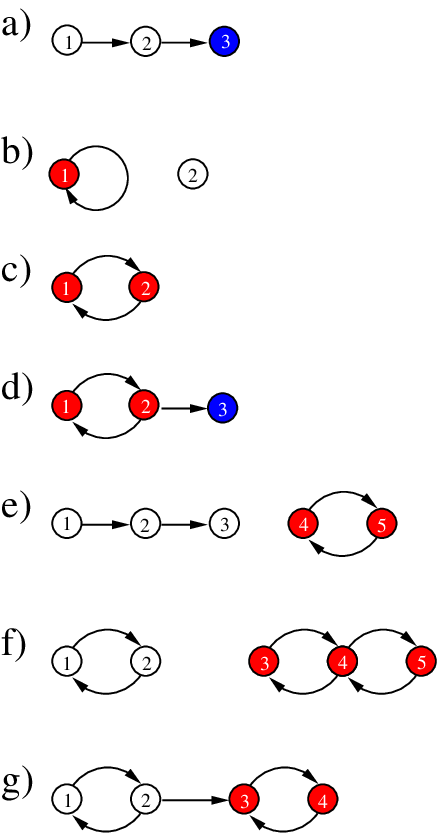}
\vchcaption{Examples of graphs with a unique PFE. The subgraph of the PFE 
coincides with the nodes that are populated in the attractor.}
\label{uniquePFEfig}
\end{vchfigure}

\begin{vchfigure}
\epsfbox{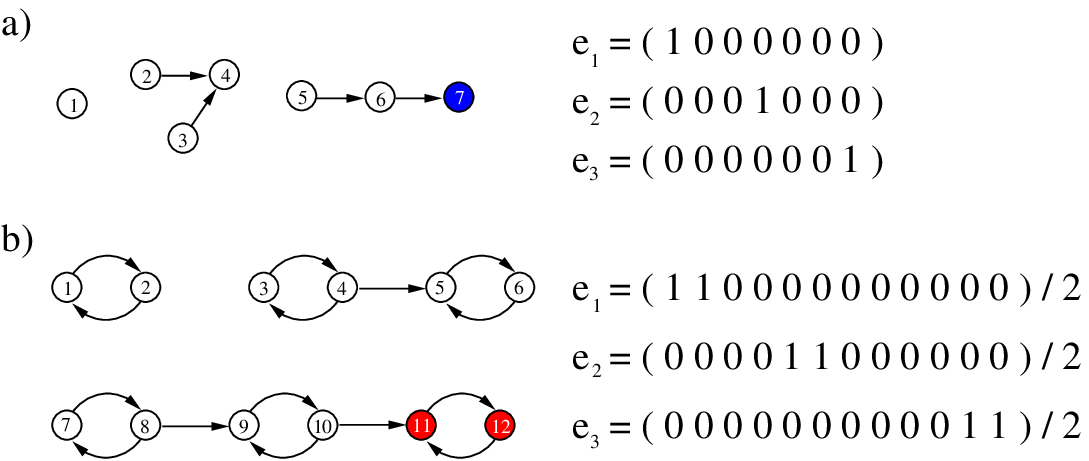}
\vchcaption{Examples of graphs with multiple PFEs. 
(a) ${\bf e}_1,{\bf e}_2,{\bf e}_3$ are all eigenvectors with eigenvalue $\lambda_1=0$.
Only ${\bf e}_3$ is the attractor. Thus for generic initial conditions, only node 7,
which sits at the end point of the longest chain of nodes is populated in the 
attractor. 
(b) ${\bf e}_1,{\bf e}_2,{\bf e}_3$ are all eigenvectors with eigenvalue $\lambda_1=1$,
but only ${\bf e}_3$ is the attractor. Only the 2-cycle of nodes 11 and 12, which
sits at the end of the longest chain of cycles, is populated in the attractor.}
\label{nonuniquePFEfig}
\end{vchfigure}

\begin{vchfigure}
\epsfbox{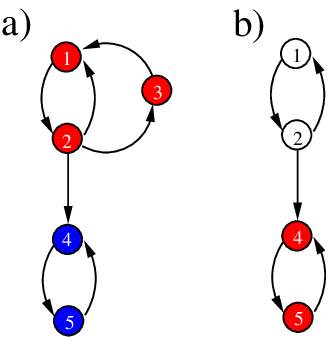}
\vchcaption{Example illustrating the notion of keystone species and
the phenomenon of a core-shift.
Node number 3 is keystone node of the graph in part {\bf a} because
its removal produces the graph in {\bf b} which has a zero core overlap with
the graph in {\bf a}. The core nodes of both graphs are coloured red.
An event in which the core before the event and after the event have
zero overlap is called a `core-shift'.}
\label{keystonefig}
\end{vchfigure}

\begin{vchfigure}
\epsfxsize=12cm
\epsfbox{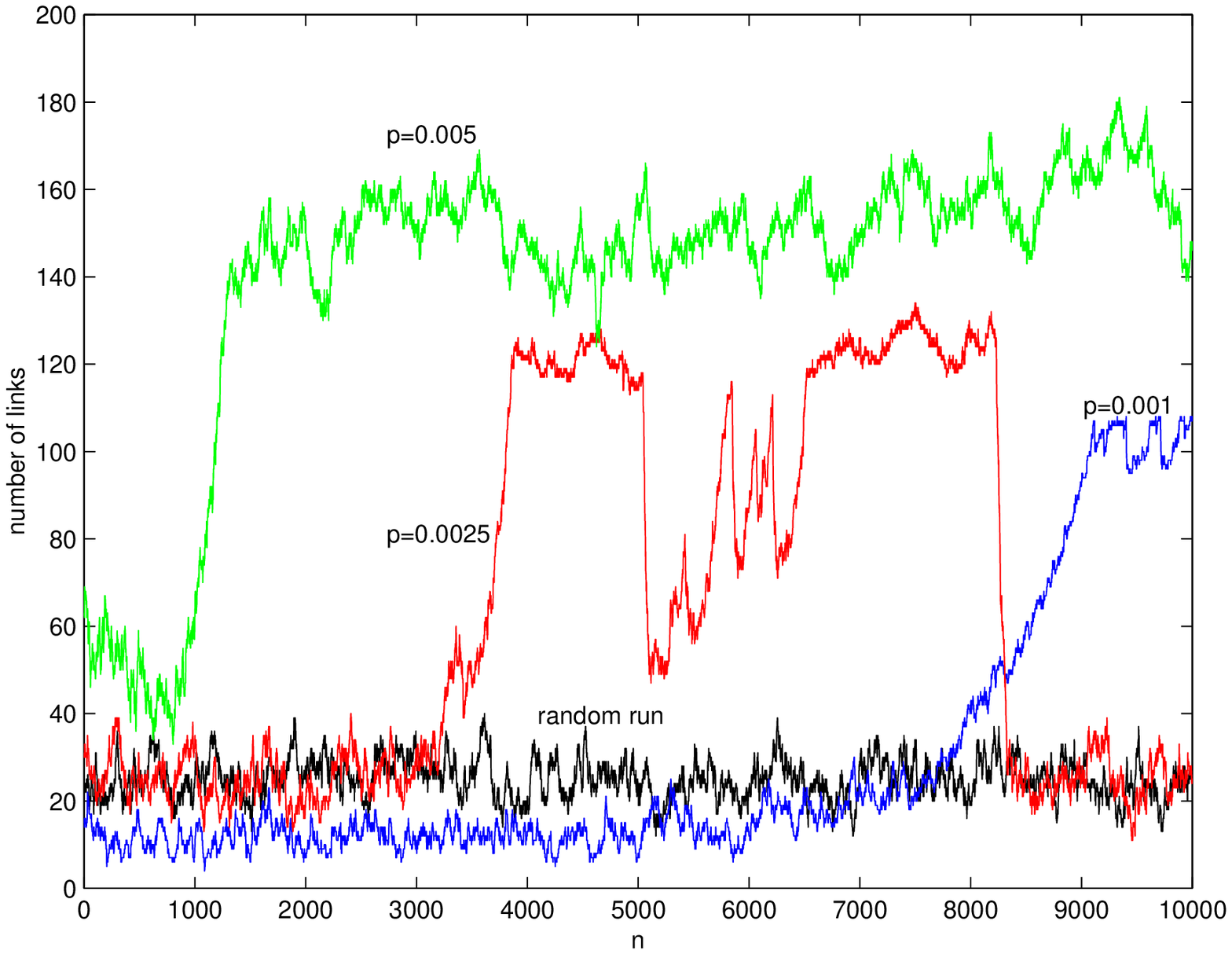}
\vchcaption{The number of links versus time (n) for various runs. Each run had
$s=100$. The black curve is a run with selection turned off; a random node
is picked for removal at each graph update. The other curves show runs with 
selection turned on and with different $p$ values: Blue $p=0.001$,
Red $p=0.0025$, Green $p=0.005$.}
\label{linksfig}
\end{vchfigure}

\begin{vchfigure}
\epsfxsize=9cm
\epsfbox{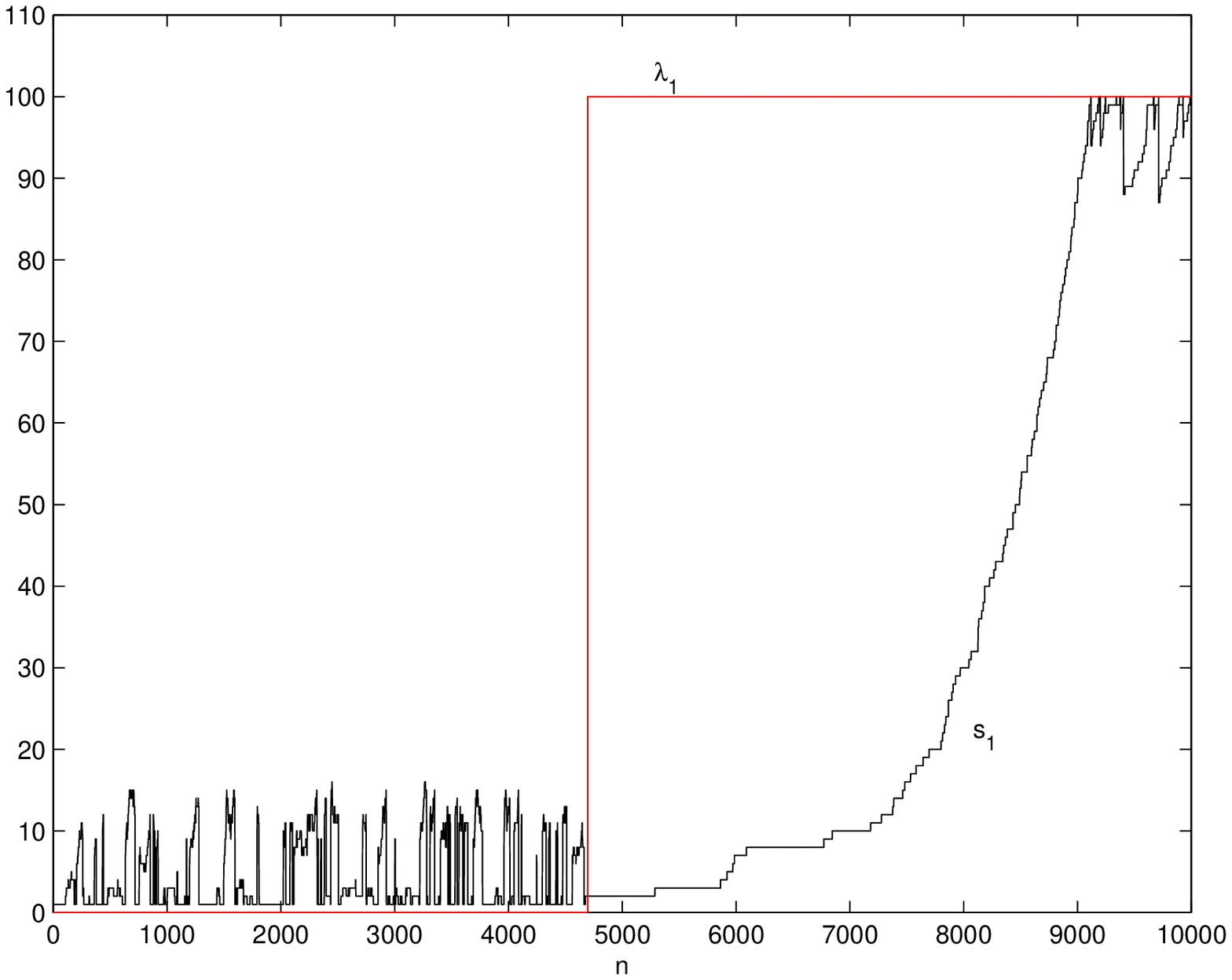}
\epsfxsize=9cm
\epsfbox{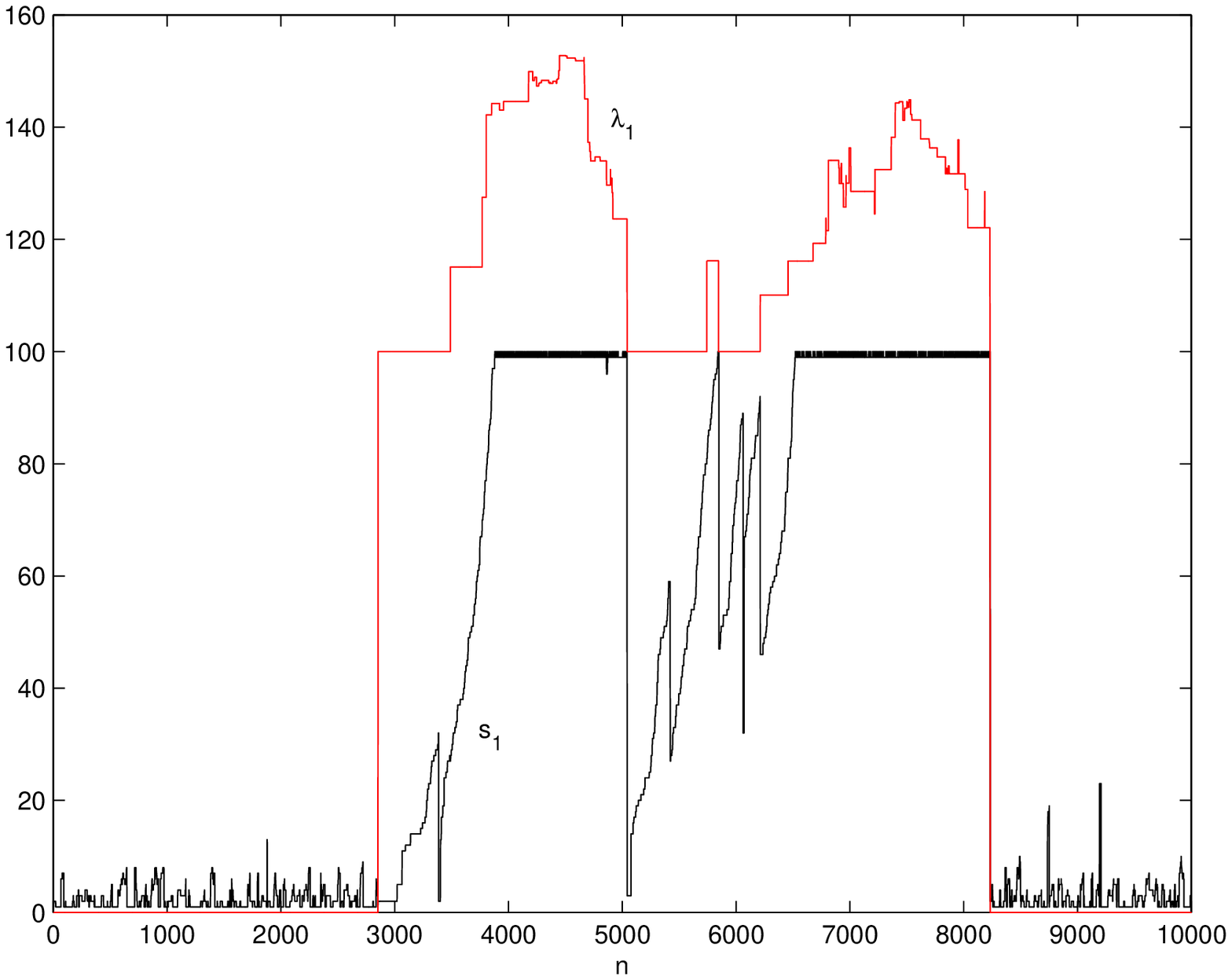}
\epsfxsize=9cm
\epsfbox{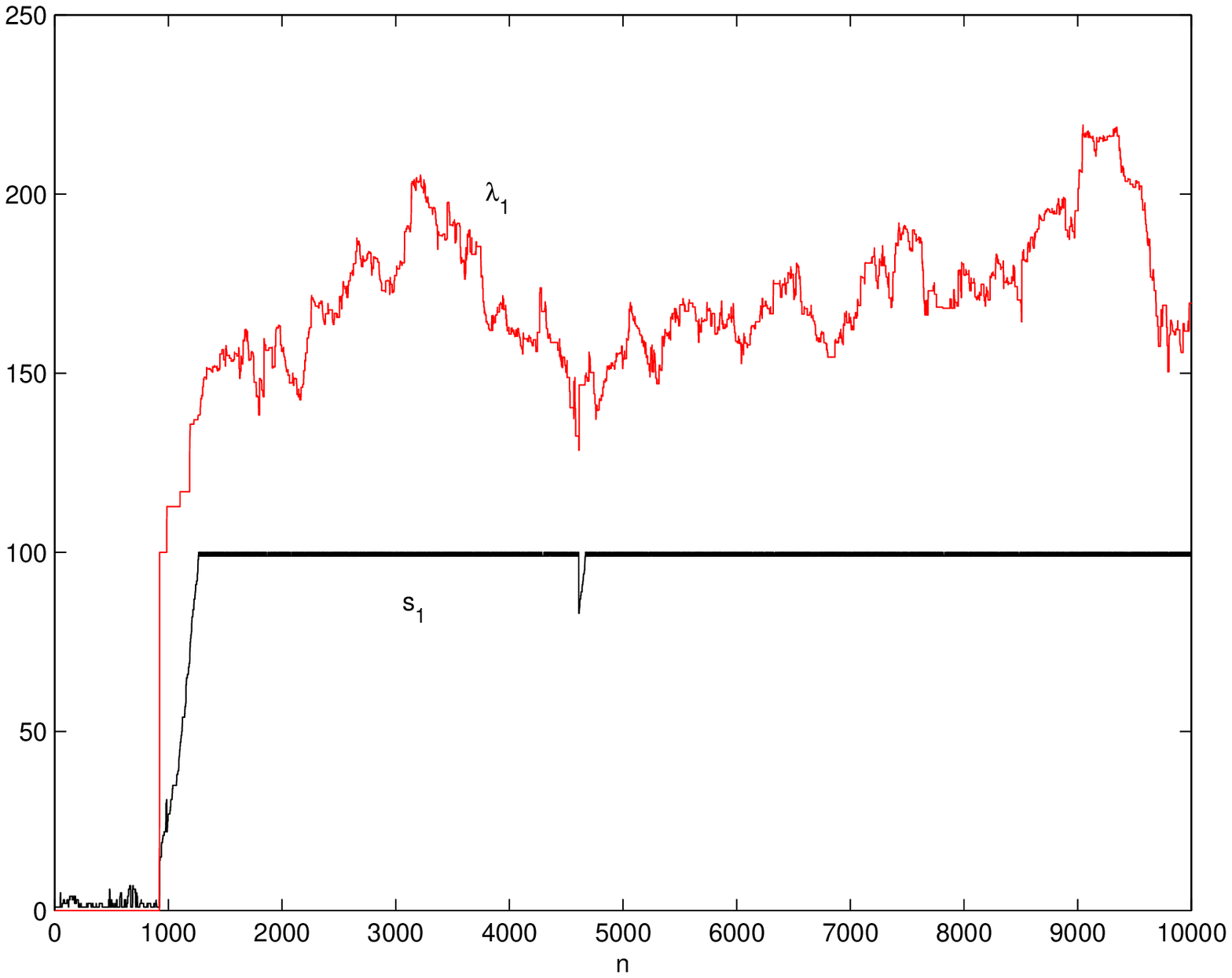}
\vchcaption{Number of populated nodes, $s_1$, (black curve) and the 
Perron-Frobenius eigenvalue of the graph, $\lambda_1$, (red curve) versus time,
$n$, for the same three runs shown in Figure \ref{linksfig}. Each run has $s=100$ and 
$p=0.001, 0.0025$ and $0.005$ respectively. The $\lambda_1$ values shown are 100
times the actual value.}
\label{s1l1fig}
\end{vchfigure}

\begin{vchfigure}
\hspace{-5cm}
\epsfbox{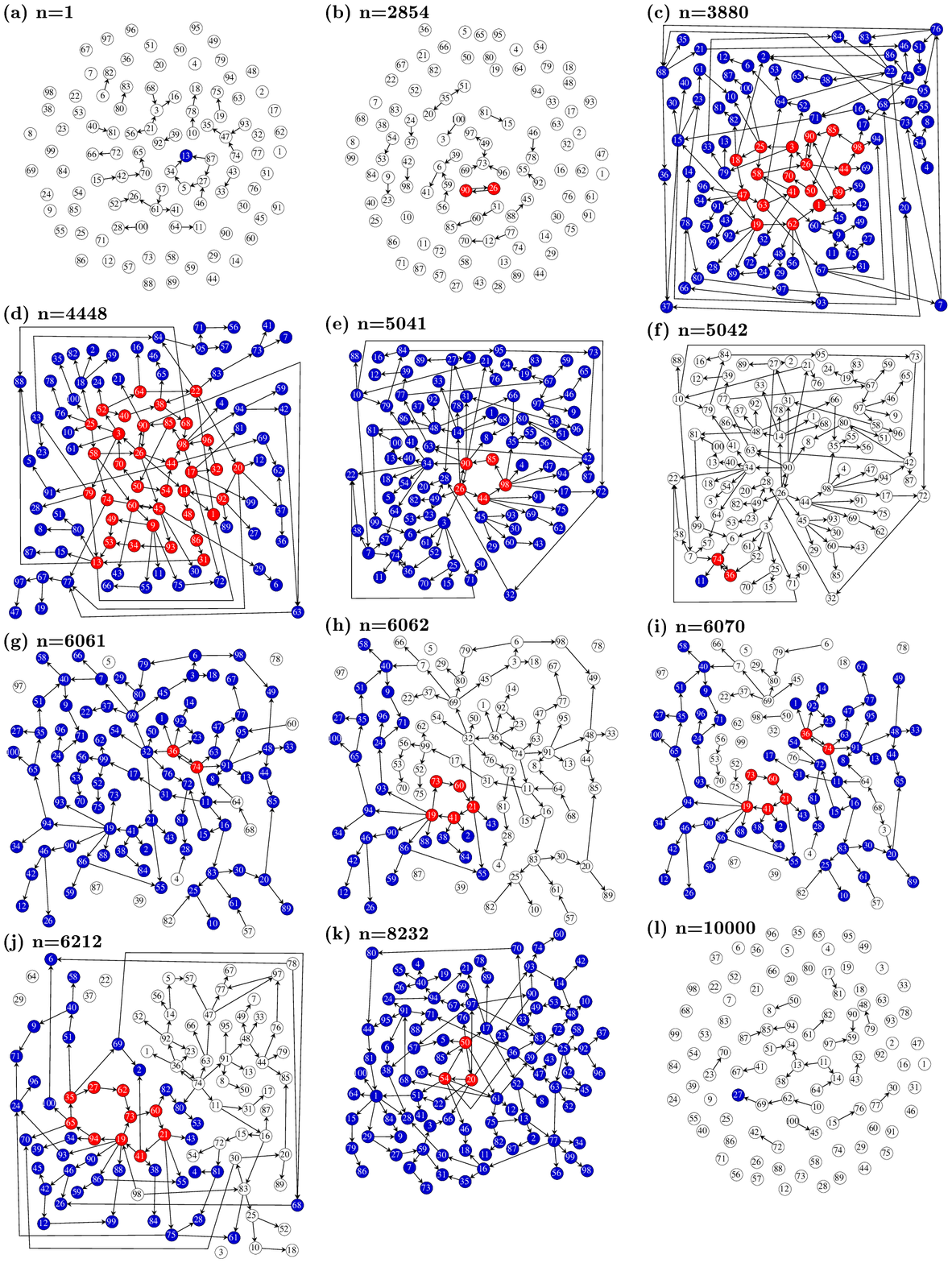}
\end{vchfigure}

\begin{vchfigure}
\vchcaption{Snapshots of the graph at various times for the run shown in Figure \ref{s1l1fig}b
with $s=100$ and $p=0.0025$. See text for a description of the major events.
In all graphs, white nodes are those with $X_i=0$. All coloured nodes have
$X_i>0$. In graphs which have an ACS, the red nodes are core nodes and the
blue nodes are periphery nodes.}
\label{snapshotsfig}
\end{vchfigure}

\begin{vchfigure}
\epsfxsize=12cm
\epsfbox{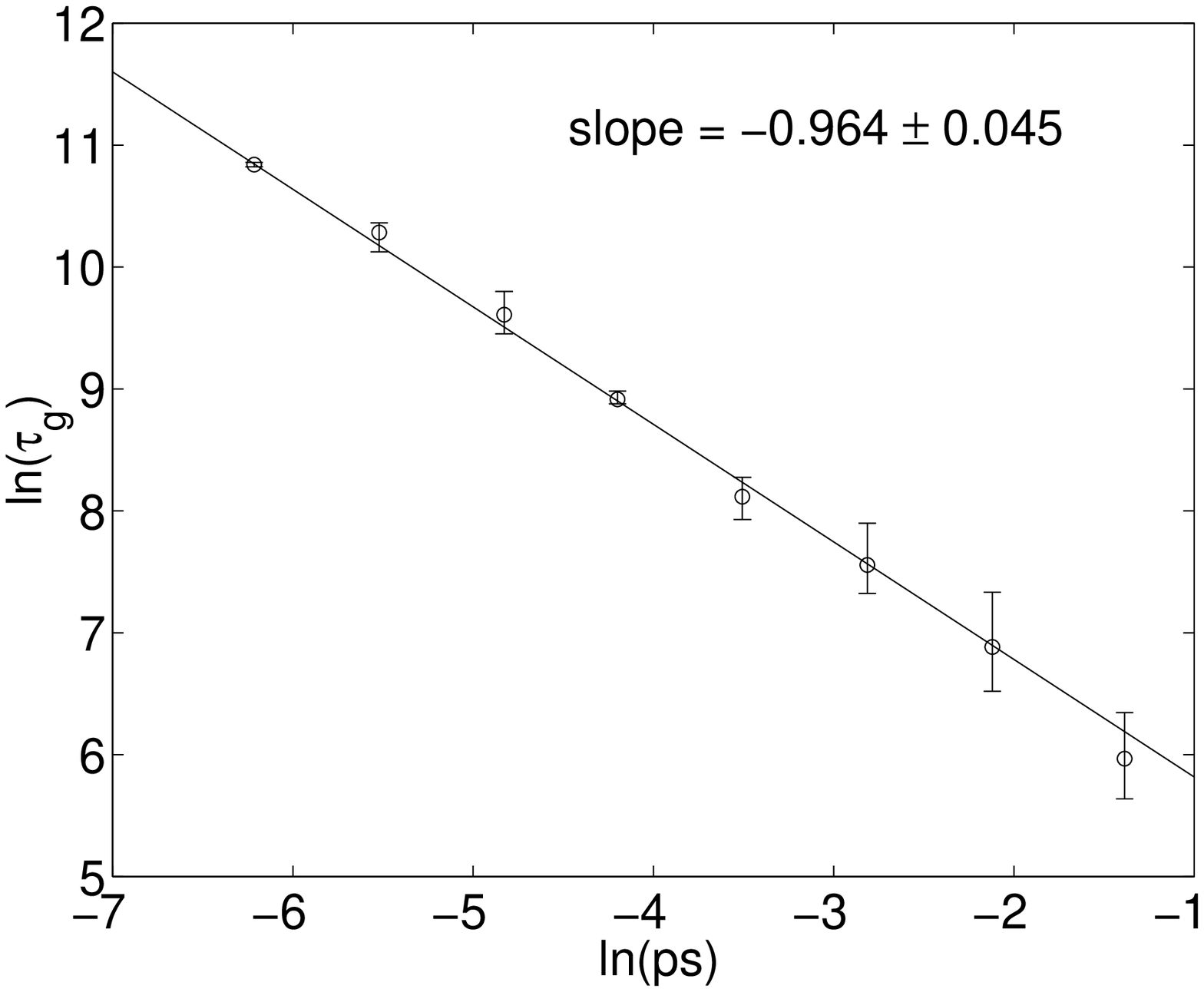}
\vchcaption{Each data point shows the average of $\tau_g$ (the growth timescale
for an ACS) over 5 different runs with $s=100$ and the given $p$ value. The error
bars correspond to one standard deviation. The solid line is the best linear fit to the data points on a log-log plot.
Its slope is consistent with the analytically predicted slope -1 (see the 
discussion of the growth phase in section 5.)}
\label{taugfig}
\end{vchfigure}

\begin{vchfigure}
\epsfxsize=12cm
\epsfbox{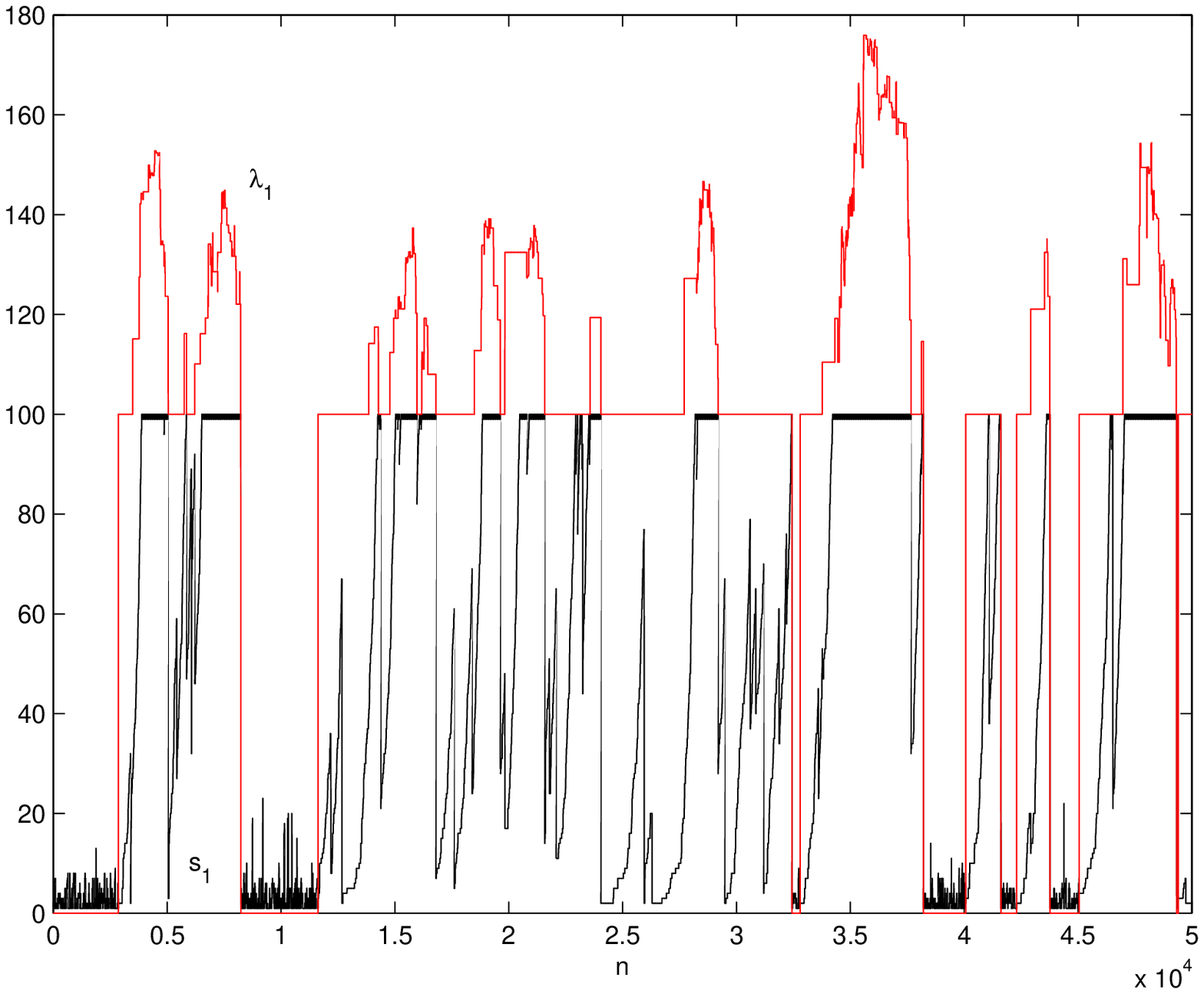}
\vchcaption{The same run displayed in Figure \ref{s1l1fig}b over a longer timescale, till 
$n=50000$. This displays repeated rounds of crashes and recoveries.}
\label{longtimefig}
\end{vchfigure}

\begin{vchfigure}
\epsfxsize=15cm
\epsfysize=20cm
\epsfbox{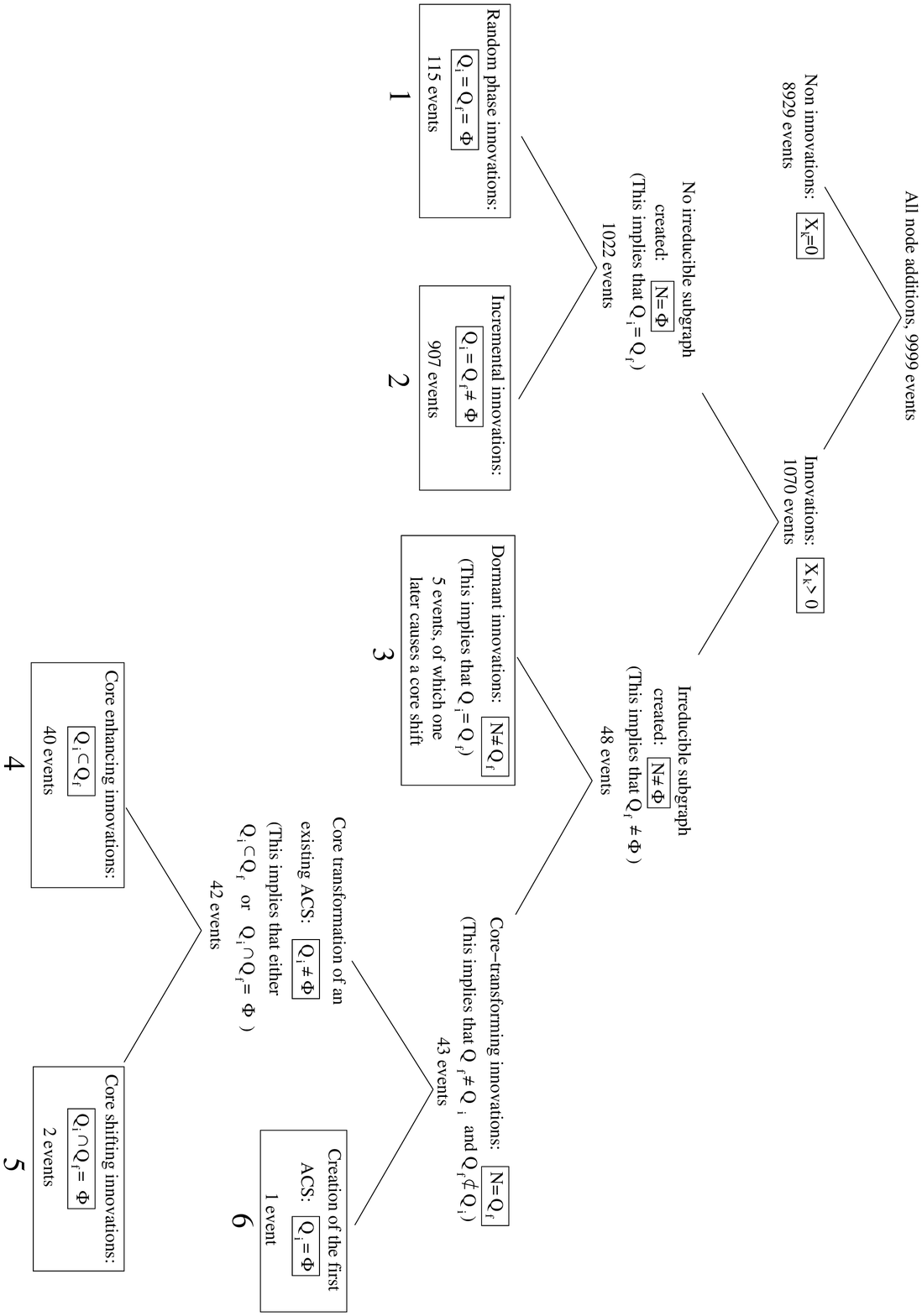}
\end{vchfigure}

\begin{vchfigure}
\vchcaption{A hierarchy of innovations. 
Each node in this binary tree represents a class of node addition events.
Each class has a name; the small box contains the mathematical
definition of the class.
All classes of events except the leaves of the tree 
are subdivided into two exhaustive and mutually exclusive 
subclasses (represented by the two branches emanating downwards from
the class). 
The number of events in each class pertain to
the run of Figure \ref{s1l1fig}b with a total of 9999 graph updates,
between $n=1$ (the initial graph) 
and $n=10000$. In that run, out of 9999 node addition events, most 
(8929 events) are not innovations. The rest (1070 events), which are
innovations, are classified according to their
graph theoretic structure. The classification is general; it is valid
for all runs. $X_k$ is the relative population of the
new node in the attractor configuration of (\ref{xdot}) that is reached
in step 1 of the dynamics (see Section 4) immediately following the 
addition of that node. $N$ stands for the new irreducible subgraph, if
any, created by the new node. If the new node causes a new irreducible 
subgraph to be created, $N$ is the {\it maximal} irreducible 
subgraph that includes the
new node. If not, $N=\Phi$ (where $\Phi$ stands for the empty set). 
$Q_i$ is the core of the graph just before the addition of the node
(just before step 3 of the dynamics in Section 4) and $Q_f$
the core just after the addition of the node. The six leaves of the
innovation subtree are numbered from 1 to 6 and correspond to the
classes discussed in Section 6. The impact of each kind of
innovation on the system dynamics is discussed in the text and in 
more detail in \cite{JK6}. Some classes of events happen
rarely (e.g., classes numbered 5 and 6) but have a major impact
on the dynamics of the system. The precise impact of all these classes of innovations
on the system over a short time scale (before the next graph
update) as well as their probable impact over the medium term
(upto a few thousand graph updates) can be predicted from the 
graph theoretic structure of $N$ and the rest of the graph at the moment
these innovations appear in a run.
}
\label{hierarchyfig}
\end{vchfigure}

\begin{vchfigure}
\epsfxsize=12cm
\epsfbox{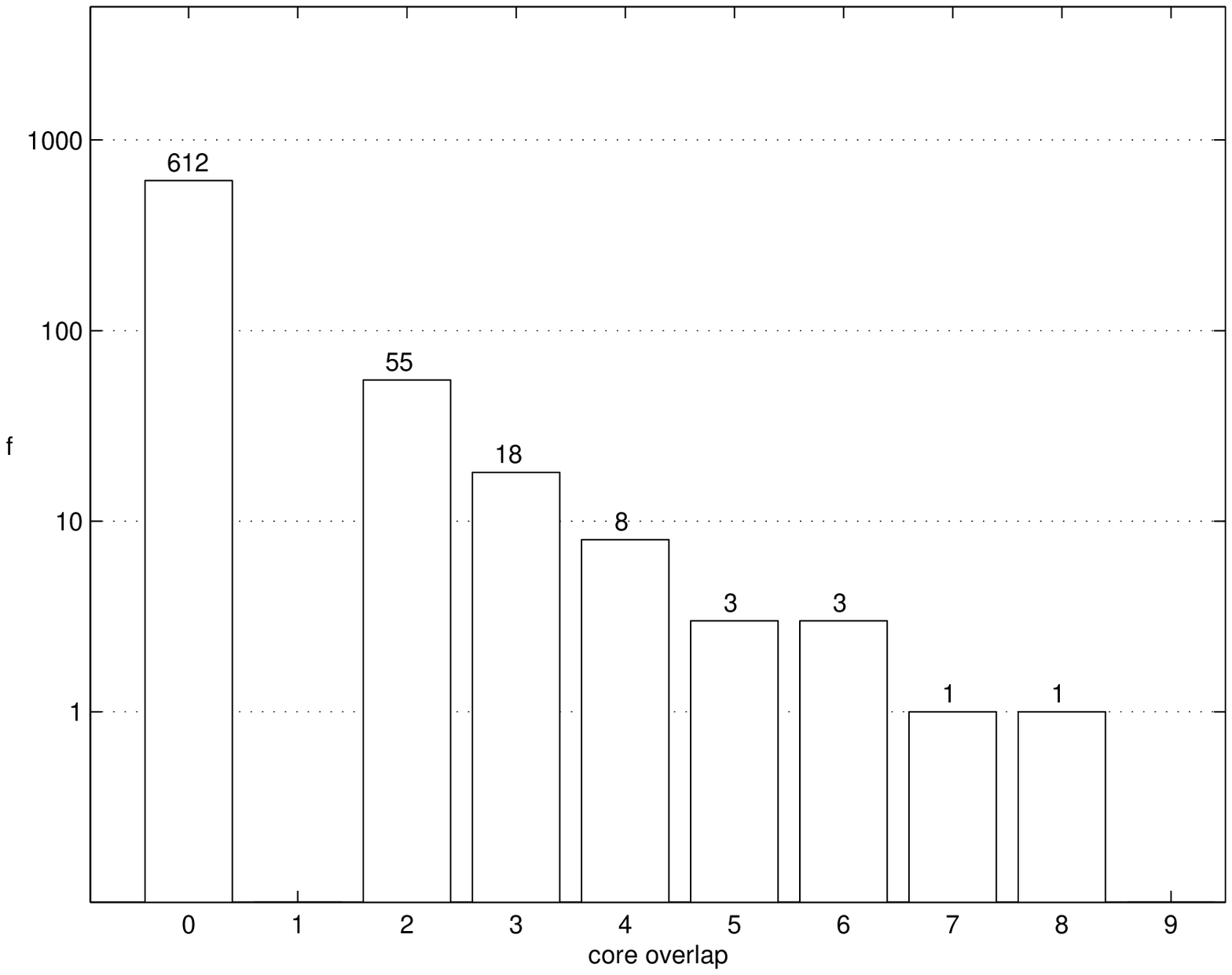}
\vchcaption{Large crashes are predominantly core-shifts.
A histogram of core overlaps for the 701 events where $s_1$ dropped
by more than $s/2$ observed in various runs with $s=100$ and $p=0.0025$,
totalling 1.55 million iterations.}
\label{histogramfig}
\end{vchfigure}

\begin{vchfigure}
\epsfxsize=12cm
\epsfbox{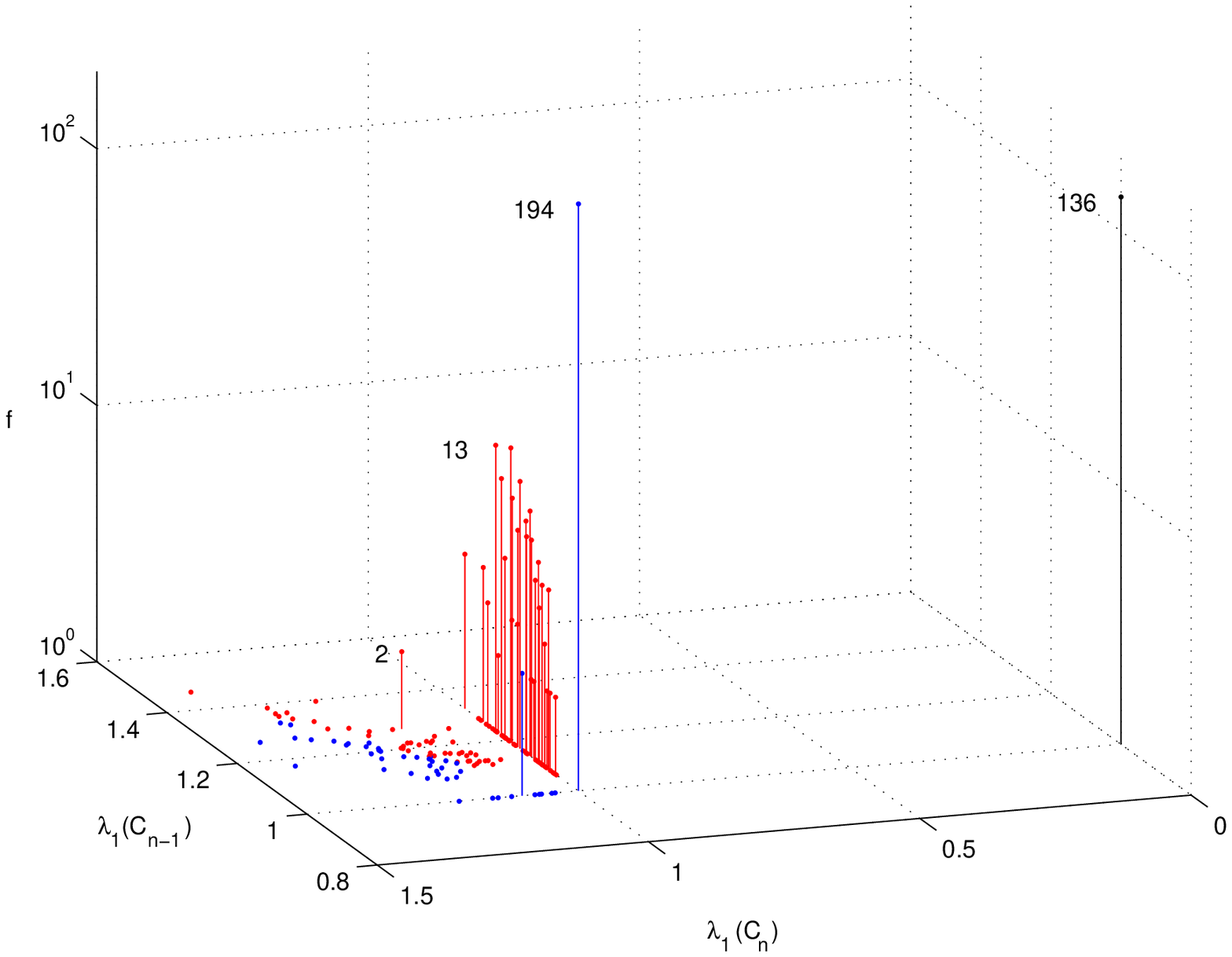}
\vchcaption{Classification of core-shifts into three categories.
The graph shows the frequency, $f$, of 
the $612$ core-shifts observed (see Figure \ref{histogramfig}) in a set of runs with $s=100$ and
$p=0.0025$ vs. the
$\lambda_1$ values before, $\lambda_1(C_{n-1})$, and after, $\lambda_1(C_n)$, the core-shift.
Complete crashes
(black; $\lambda_1(C_{n-1})=1$, $\lambda_1(C_n)=0$),
takeovers by core-transforming innovations (blue; $\lambda_1(C_n)\ge\lambda_1(C_{n-1})\ge 1$)
and takeovers by dormant innovations (red; $\lambda_1(C_{n-1})>\lambda_1(C_n)\ge 1$) are
distinguished. Numbers alongside vertical lines represent the corresponding $f$
value.}
\label{classifig}
\end{vchfigure}

\begin{thebibliography}{99}

\bibitem{AB}{Albert, R. and Barabasi, A.-L. (2002)
Statistical mechanics of complex networks,
{\it Rev. Mod. Phys.} {\bf 74}, 47
{\it (www.arXiv.org/abs/cond-mat/0106096)}.}

\bibitem{DM}{Dorogovtsev, S. N. and Mendes, J. F. F. (2002)
Evolution of networks, {\it Adv. Phys.} {\bf 51} 1079
{\it (www.arXiv.org/abs/cond-mat/0106144)}.}

\bibitem{Strogatz}{Strogatz, S. H. (2001)
Exploring complex networks, {\it Nature} {\bf 410}, 268-276.}

\bibitem{Watts}{Watts, D. J. (1999) {\it Small Worlds: The dynamics of Networks 
between Order and Randomness} (Princeton Univ. Press, Princeton).}

\bibitem{Dyson}{Dyson, F. (1985) {\it Origins of Life} (Cambridge Univ. Press
Cambridge, UK).}

\bibitem{FKP}
{Farmer, J. D., Kauffman, S. and Packard, N. H. (1986) 
Autocatalytic replication of polymers,
{\it Physica}
{\bf D22} 50-67.}

\bibitem{BFF}
{Bagley, R. J., Farmer, J. D. and Fontana, W. (1991) 
Evolution of a metabolism,
in {\it Artificial Life II}, eds. Langton, C. G., Taylor, C., Farmer, J. D.
and Rasmussen, S.
(Addison Wesley, Redwood City),
pp. 141-158.}

\bibitem{Kauffman2}
{Kauffman, S. A. (1993)
{\it The Origins of Order}
(Oxford Univ. Press).}

\bibitem{BS}
{Bak, P. and Sneppen, K. (1993)
Punctuated equilibrium and criticality in a simple
model of evolution,
{\it Phys. Rev. Lett.} {\bf 71}, 4083-4086.}

\bibitem{FB}{Fontana, W. and Buss, L. (1994)
The arrival of the fittest: Toward a theory
of biological organization,
{\it Bull. Math. Biol.} {\bf 56}, 1-64.}

\bibitem{Harary}{Harary, F. (1969) {\it  Graph Theory} (Addison Wesley, Reading, MA, USA).}     

\bibitem{BG}{Bang-Jensen, J. and Gutin, G. (2001) {\it Digraphs: Theory,
Algorithms and Applications} (Springer-Verlag, London).}

\bibitem{Seneta}
{Seneta, E. (1973) {\it Non-Negative Matrices} (George Allen
and Unwin, London).}

\bibitem{BP}{Berman, A. and Plemmons, R. J. (1994)
{\it Non-negative matrices in the mathematical sciences} (SIAM, Philadelphia).}

\bibitem{Rothblum}{Rothblum, U. G. (1975)
Algebraic eigenspaces of nonnegative matrices, {\it Linear Algebra and Appl}
{\bf 12}, 281-292.}

\bibitem{JK2}
{Jain, S. and Krishna, S. (1999) 
Emergence and growth of complex networks in adaptive
systems,
{\it Computer Physics Comm.} {\bf 121-122}, 116-121.}

\bibitem{Eigen}{Eigen, M. (1971)
Self-organization of matter and the evolution of
biological macromolecules,
{\it Naturwissenschaften} {\bf 58}, 465-523.}

\bibitem{Kauffman1}{Kauffman, S.A. (1971)
Cellular homeostasis, epigenesis and
replication in randomly aggregated macromolecular systems,
{\it J. Cybernetics} {\bf 1}, 71-96.}

\bibitem{Rossler}{Rossler, O. E. (1971)
A system theoretic model of biogenesis,
{\it Z. Naturforschung} {\bf 26b}, 741-746.}

\bibitem{JK1}
{Jain, S. and Krishna, S. (1998)
Autocatalytic sets and the growth of complexity in
an evolutionary model,
{\it Phys. Rev. Lett.} {\bf 81}, 5684-5687.}

\bibitem{JK3}
{Jain, S. and Krishna, S. (2001) 
A model for the emergence of cooperation,
interdependence and structure in evolving networks,
{\it Proc. Natl. Acad. Sci. (USA)} {\bf 98}, 543-547.} 

\bibitem{JK4}
{Jain, S. and Krishna, S. (2002) 
Crashes, recoveries and `core-shifts' in a model of evolving
networks,
{\it Phys. Rev. E} {\bf 65}, 026103,
{\it www.arXiv.org/abs/nlin.AO/0107037}.}

\bibitem{JK5}
{Jain, S. and Krishna, S. (2001) 
Large extinctions in an evolutionary model:
the role of innovation and keystone species,
{\it Proc. Natl. Acad. Sci. (USA)} 
{\bf 99}, 2055-2060,
{\it www.arXiv.org/abs/nlin.AO/0107038}.}

\bibitem{Paine}
{Paine, R. T. (1969) 
A note on trophic complexity and community stability,
{\it Am. Nat.}
{\bf 103}, 91-93.}

\bibitem{Pimm}{Pimm, S. L. (1991) {\it The Balance of Nature?
Ecological Issues in the Conservation of Species and Communities} 
(Univ. of Chicago Press, Chicago).}

\bibitem{JTM}
{Jord\'{a}n, F., Tak\'{a}cks-S\'{a}nta, A. and Moln\'{a}r, I. (1999)
A reliability theoretical quest
for keystones,
{\it OIKOS} {\bf 86}, 453-462.}

\bibitem{SMo} {Sol\'{e}, R. V. and Montoya, J. M. (2000) 
Complexity and fragility in
ecological networks,
{\it www.arXiv.org/abs/cond-mat/0011196}.}

\bibitem{JSMO}{Joyce, G. F., Schwartz, A. W., Miller, S. L. and Orgel, L. E. (1987)
The case for an ancestral genetic system involving simple analogues of the nucleotides,
{\it Proc. Natl. Acad. Sci. (USA)} {\bf 84}, 4398-4402.}

\bibitem{Joyce}{Joyce, G. F. (1989) RNA evolution and the origins of life,
{\it Nature} {\bf 338}, 217-223.}

\bibitem{JK6}{Jain, S. and Krishna, S. (2002)
Constructive and destructive effects of `innovation' in evolving networks,
Preprint 2002.}

\bibitem{MKYC}{Morowitz, H. J., Kostelnik, J. D., Yang, J. and
Cody, G. D. (2000) The origin of intermediary metabolism,
{\it Proc. Natl. Acad. Sci. (USA)} {\bf 97}, 7704-7708.}


\bibitem{SBDL}{Segr\'{e}, D., Ben-Eli, D., Deamer, W. D. and Lancet, D. (2001)
The lipid world, {\it Origins of Life and Evol. of the Biosphere} {\bf 31}, 119-145.}


\end{thebibliography}
\end{document}